\documentclass[11pt]{article}
\usepackage{amsmath, amssymb}
\usepackage{graphicx}
\usepackage{natbib}
\usepackage{url} 
\usepackage{color, colortbl}
\usepackage{geometry}
\usepackage{array}
\usepackage{caption}
\usepackage{subcaption}
\usepackage{mdframed}
\usepackage{setspace}
\usepackage{hyperref}
\hypersetup{
	colorlinks= true,
	linkcolor= cyan,
	citecolor= blue,
	urlcolor= blue,
}

\newmdtheoremenv{prop}{Theorem}
\newmdtheoremenv{lemma}{Lemma}

\newcommand{\btheta}{\boldsymbol{\theta}}

\newcommand{\bgamma}{\boldsymbol{\gamma}}
\newcommand{\bxi}{\boldsymbol{\xi}}
\newcommand{\bfeta}{\boldsymbol{\eta}}
\newcommand{\bfomega}{\boldsymbol{\omega}}

\definecolor{lightgray}{gray}{0.9} 

\title{Gibbs sampling for Bayesian P-splines}
\author{Oswaldo Gressani$^{1\star}$ and Paul H.C.\ \hspace{-0.2cm} Eilers$^{2}$\\}
\date{}

\begin{document}
	
\maketitle 

\noindent $^{1}$ Interuniversity Institute for Biostatistics and statistical Bioinformatics (I-BioStat), Data Science Institute, Hasselt University, Hasselt, Belgium.\\
\\
\noindent $^{2}$ Department of Biostatistics, Erasmus University Medical Center, Rotterdam, The Netherlands.\\

\vspace{2cm}

\begin{center}
\Large{\textbf{Abstract}}
\end{center}

\noindent P-splines provide a flexible setting for modeling nonlinear model components based on a discretized penalty structure with a relatively simple computational backbone. Under a Bayesian inferential framework based on Markov chain Monte Carlo, estimates of model coefficients in P-splines models are typically obtained by means of Metropolis-type algorithms. These algorithms rely on a proposal distribution that has to be carefully chosen to generate Markov chains that efficiently explore the parameter space. To avoid such a sensitive tuning choice, we extend the Gibbs sampler to Bayesian P-splines models. In this model class, conditional posterior distributions of model coefficients are shown to have attractive mathematical properties. Taking advantage of these properties, we propose to sample the conditional posteriors by alternating between the adaptive rejection sampler when targets are log-concave and the Griddy-Gibbs sampler when targets are characterized by more complex shapes. The proposed Gibbs sampler for Bayesian P-splines (GSBPS) algorithm is shown to be an interesting tuning-free tool for inference in Bayesian P-splines models. Moreover, the GSBPS algorithm can be translated in a compact and user-friendly routine. After describing theoretical results, we illustrate the potential of our methodology in density estimation, Binomial regression, and smoothing of epidemic curves.\\


\vspace{5cm}

\noindent {\small{$^\star$ Corresponding author: oswaldo.gressani@uhasselt.be}}

\newpage 

\section{Introduction}
	
\noindent The popularity of P-splines \citep{eilers1996flexible} owes much to the simple and elegant idea of combining many B-spline basis functions with a discrete roughness penalty imposed on their coefficients. P-splines are rooted in generalized linear models and endowed with appealing computational properties. This makes them a particularly attractive smoother with wide-ranging applications. Another important feature of P-splines is that they can be formulated in a Bayesian framework \citep{lang2004bayesian}. Under the Bayesian paradigm, B-spline coefficients and the penalty parameter can be estimated jointly by exploring the posterior distribution obtained from Bayes' theorem. Frequentist statistical procedures relying on the penalized likelihood principle do not benefit from such a privilege in general, as the penalty parameter undergoes a separate analysis with cross-validation methods or minimization of a chosen information criterion. A Bayesian approach for P-splines naturally deals with the problem of selecting an appropriate amount of smoothness in a data-driven way and, as such, is a convincing argument for users in demand for an automated selection of the penalty parameter.\\
\indent In general, Bayesian P-splines models rely on Markov chain Monte Carlo (MCMC) methods to generate samples from posterior distributions and carry out inference. When Markov chain simulation becomes computationally demanding, sampling-free approaches involving Laplace approximations (Laplacian-P-splines) provide an interesting alternative \citep[see e.g.,][]{gressani2018fast,  gressani2021laplace, gressani2022laplacian, lambert2023penalty}. MCMC algorithms in Bayesian P-splines models traditionally sample the posterior by means of a Metropolis-within-Gibbs strategy, with a Gibbs step for the penalty parameter and other hyperparameters in situations of conjugacy \citep{bremhorst2016flexible, lambert2009bayesian, gressani2022epilps}. B-spline coefficients and other regression coefficients are sampled with a Metropolis-Hastings algorithm \citep{metropolis1953equation, hastings1970monte} or a Metropolis-adjusted Langevin algorithm \citep{roberts1996exponential}. The efficiency with which Markov chains generated by Metropolis-type algorithms explore the posterior target distribution heavily hinges on the chosen proposal distribution. The degree of difficulty of such a key choice increases with the dimension of the space to be sampled. Manual tuning of the proposal by trial and error can rapidly become a hopeless task even under a parameter space with a moderate number of dimensions. This rather challenging problem is lurking in many models and Bayesian P-splines are no exception. Adaptive MCMC \citep[e.g.][]{haario2001adaptive, roberts2009examples} offers a possibility to circumvent this at the cost of additional programming steps. A more suitable alternative in the context of Bayesian P-splines would be a MCMC procedure that entirely avoids the choice of a proposal distribution. Gibbs sampling is such a procedure. The Gibbs sampler was pioneered by \cite{geman1984stochastic} in the context of image processing and sparked a revolution in computational Bayesian methods \citep{brooks2011handbook} thanks to the papers of \cite{gelfand1990sampling} and \cite{casella1992explaining}. To our knowledge, no attempts have yet been made to explore the joint posterior distribution in Bayesian P-splines models by entirely relying on the Gibbs sampler when data are non-normal. The reason for such a gap in the literature is fairly simple. In non-normal situations, the classic Gaussian random walk priors imposed on B-spline coefficients \citep{lang2004bayesian} are not conditionally conjugate and conditional posterior distributions have a non-standard form. In many cases, other model parameters (e.g.\ additional regression parameters or hyperparameters) also do not meet the conjugacy property. As such, a MCMC algorithm entirely relying on the Gibbs sampler in Bayesian P-splines models appears to be beyond reach.\\
\indent In this article, we argue that this is not true. In fact, under certain regularity conditions that are easy to check in practice, we show that the Gibbs sampler is a viable option to explore the joint posterior distribution of the entire set of parameters in Bayesian P-splines models. Working under the popular class of Gaussian random walk priors for the B-spline coefficients \citep{lang2004bayesian}, we derive a closed-form expression for the conditional priors that is easy to evaluate and that can in turn be used to efficiently evaluate the conditional posteriors of the B-spline coefficients. Our Gibbs sampling scheme alternates between adaptive rejection sampling \citep{gilks1992adaptive} for (univariate) log-concave conditional posterior targets and Griddy-Gibbs sampling \citep{ritter1992facilitating} for (univariate) conditional posterior targets that are not log-concave. Griddy-Gibbs sampling is a tool to draw samples from univariate conditional posterior distributions based on a discrete approximation of conditional posterior cumulative distribution functions.  It is particularly attractive for three main reasons: (1) it serves as a surrogate to the Gibbs sampler when models are not conditionally conjugate; (2) its algorithmic implementation is rather simple; and (3) grids on which to evaluate the conditionals can be flexibly adapted depending on the targeted approximation accuracy.\\
\indent The article is organized as follows. In Section 2, we present the Bayesian P-splines model. The choice of penalty orders and priors is also discussed here. This section also contains core results related to univariate conditional posterior distributions of model coefficients. Section 3 gives a detailed exposition of the Gibbs sampler for Bayesian P-splines (GSBPS) algorithm along with a pseudo-code. Section 4 covers real applications in density estimation, binomial regression and smoothing of epidemic curves. Finally, Section 5 concludes by highlighting the strengths and limitations of the proposed method. Results in this paper can be reproduced with the codes provided in the following repository \url{https://github.com/oswaldogressani/GSBPS}.

\section{Bayesian P-splines}

\subsection{General formulation}

\noindent Parameters in P-splines models can generally be separated in three distinct categories. The first category includes the spline coefficients and we denote by $\btheta=(\theta_1,\dots,\theta_K)^{\top}$ the vector of B-spline coefficients associated to $K$ B-spline basis functions with $K \in \mathbb{N}:=\{1,2,3,\dots\}$. The ``P'' in P-splines stands for penalties, meaning that the parameters of the first category are penalized. The second category gathers additional parameters that are not penalized (e.g.\ regression coefficients) and we use the notation $\bgamma=(\gamma_1,\dots,\gamma_l)^{\top}$ for the vector including $l$ unpenalized parameters. The third category includes the hyperparameters (e.g.\ the penalty parameter) of the model and $\bfeta=(\eta_1,\dots,\eta_p)^{\top}$ represents a vector of $p$ hyperparameters. Parameters in the first and second category are usually labeled as latent parameters and gathered in a single vector notation $\bxi=(\btheta^{\top}, \bgamma^{\top})^{\top}$ to easily distinguish them from the hyperparameters. In its most simple form, a P-splines model will include the first and third category and the latter will only contain parameters related to the penalty. The compact notation $\bfomega=(\btheta^{\top}, \bgamma^{\top}, \bfeta^{\top})^{\top}$ denotes the vector including all $J=K+l+p$ model parameters and $\omega_j$ is used to generically denote the $j$th component of $\bfomega$. 

\subsection{Priors}

\noindent Working under the Bayesian paradigm, we specify our degree of belief on the model parameters by means of prior distributions. A commonly used prior for the B-spline coefficients is the so-called global smoothness prior of \cite{lang2004bayesian} based on Gaussian random walks $\btheta \vert \lambda \sim \mathcal{N}(\boldsymbol{0}, \Sigma_{\lambda})$, with covariance matrix $\Sigma_{\lambda}=(\lambda P)^{-1}$. The covariance matrix includes the penalty parameter $\lambda\in \mathbb{R}_+:=\{x \in \mathbb{R} \vert x>0\}$ responsible for tuning smoothness and the penalty matrix $P=D_r^{\top}D_r+\varepsilon I_K$, where $r \in \mathbb{N}$ is the penalty order and $D_r$ is the $r$th order difference matrix of dimension $(K-r) \times K$. Here, we focus on penalties of order two ($r=2$) and three ($r=3$) since higher orders are rarely useful in practice. The diagonal perturbation with identity matrix $I_K$ and  $\varepsilon>0$ ensures $P$ is full rank, i.e.\ $\text{rank}(P)=K$. Following previous work, we fix $\varepsilon=10^{-6}$ \citep[see e.g.][]{cetinyurek2011smooth, lambert2020inclusion}. The difference matrix for a second-order and third-order penalty, respectively, takes the following form:

\begin{eqnarray}
	D_2=\begin{pmatrix}
		\phantom{-}1 & -2 & \phantom{-}1 & \phantom{-}0 & \dots & \phantom{-}0 \\
		\phantom{-}0 & \phantom{-}1 & -2 & \phantom{-}1 & \dots & \phantom{-}0 \\
		\vdots & \ddots & \ddots & \ddots & \ddots & \vdots \\
		\phantom{-}0 & \phantom{-}0 & \dots & \phantom{-}1 & -2 & \phantom{-}1
	\end{pmatrix}, \hspace{1cm} D_3=\begin{pmatrix}
		-1 & \phantom{-}3 & -3 & \phantom{-}1 & 0 & \dots & 0 \\
		\phantom{-}0 & -1 & \phantom{-}3 & -3 & \phantom{-}1 & \dots & 0 \\
		\vdots &  \ddots & \ddots & \ddots & \ddots & \ddots & \vdots \\
		\phantom{-}0 & \phantom{-}0 & \dots & -1 & \phantom{-}3 & -3 & \phantom{-}1
	\end{pmatrix}. \nonumber 
\end{eqnarray}

\vspace{0.25cm}

\noindent The prior specification for the penalty parameter is usually articulated around a Gamma distribution, as originally proposed in \cite{lang2004bayesian}. Their suggestion of a simple Gamma prior $\lambda \sim \mathcal{G}(a,b)$ (with shape $a>0$ and rate $b>0$) comes with a warning regarding the sensitivity of the estimated spline coefficients with respect to the selected hyperparameters $a$ and $b$. A more robust prior specification that avoids such a sensitive choice of hyperparameters is proposed by \cite{jullion2007robust}. They recommend to replace the shape parameter by $\nu/2$ (with $\nu>0$) and rate parameter by $\nu \delta/2$, combined with the hyperprior $\delta \sim \mathcal{G}(a_{\delta}, b_{\delta})$, and show that the resulting spline estimates are insensitive to the values selected for $\nu$, $a_{\delta}$ and $b_{\delta}$. A default choice for the latter parameters is usually $a_{\delta}=b_{\delta}=10^{-4}$ and $\nu=2$ \citep[see e.g.][]{bremhorst2016flexible, lambert2020inclusion}. The special case $a_{\delta}=b_{\delta}=0.5$ and $\nu=1$ yields a half-Cauchy prior for $\sqrt{\lambda}$ with $p(\lambda)\propto \lambda^{-0.5} (1+\lambda)^{-1}$ \citep{lambert2019estimation}. The robustness brought by the \cite{jullion2007robust} prior motivates us to work with $\lambda \vert \delta \sim \mathcal{G}(\nu/2, \nu \delta/2)$ and $\delta \sim \mathcal{G}(a_{\delta},b_{\delta})$. Priors imposed on other hyperparameters in $\bfeta$ and on non-penalized parameters in $\bgamma$ will depend on the underlying model and on the available prior information. A Gaussian prior is usually imposed on non-penalized parameters $\bgamma \sim \mathcal{N}(\mu_{\bgamma},\Sigma_{\bgamma})$ \citep{lambert2023penalty, gressani2022laplacian, gressani2021laplace} resulting in a global Gaussian prior for the latent vector $\bxi$.

\subsection{Conditional posteriors}

\noindent Using Bayes' theorem, the joint posterior distribution is given by $p(\bfomega\vert \mathcal{D}) \propto \mathcal{L}(\bfomega; \mathcal{D}) p(\bfomega)$, where $\mathcal{D}$ represents the observable units or data, the proportionality symbol $\propto$ denotes equality up to a multiplicative constant, $\mathcal{L}(\bfomega; \mathcal{D})$ is the likelihood function (asumed to be twice differentiable with respect to its arguments), and $p(\bfomega)$ represents the prior distribution described in Section 2.2. The univariate conditional posterior for component $\omega_j$ of $\bfomega$ can be written as $p(\omega_j \vert \bfomega_{-j}, \mathcal{D}) \propto \mathcal{L}(\omega_j; \bfomega_{-j}, \mathcal{D})p(\omega_j \vert \bfomega_{-j}) p(\bfomega_{-j})$, where $\bfomega_{-j}$ is vector $\bfomega$ without the $j$th component and $\mathcal{L}(\omega_j; \bfomega_{-j}, \mathcal{D})$ is the conditional likelihood seen as a function of $\omega_j$ with all remaining model parameters fixed. In general, most model parameters $\omega_j$ will have a conditional posterior that does not belong to a classic family of distributions. Therefore, construction of a sampling scheme to explore such irregular conditionals becomes challenging and the Gibbs sampling strategy fades away. In this section, we show that the classic prior structure assumed in Section 2.2 yields a (strictly) concave univariate conditional posterior $p(\omega_j \vert \bfomega_{-j}, \mathcal{D})$ for a B-spline coefficient $\omega_j$ provided that the conditional likelihood function is concave with respect to $\omega_j$. This is also true when $\omega_j$ belongs to the category of unpenalized model parameters. As such, when Bayesian P-splines models satisfy this condition, it is possible to set up an efficient Gibbs sampling scheme to explore the joint posterior by leveraging the adaptive rejection sampler of \cite{gilks1992adaptive}. For model parameters having a conditional likelihood that is not concave with respect to its argument or when it is mathematically difficult to assess the concavity property, we suggest to rely on the Griddy-Gibbs sampling procedure proposed by \cite{ritter1992facilitating}.

\subsubsection{Conditional posteriors of $\lambda$ and $\delta$}

\noindent The conditional prior of the penalty parameter is $p(\lambda \vert \delta) \propto \lambda^{0.5\nu-1}\exp(-0.5\lambda \nu \delta)$ and the conditional posterior is $p(\lambda \vert \bfomega_{-j}, \mathcal{D})\propto p(\lambda \vert \delta) p(\btheta \vert \lambda)$, with $p(\btheta \vert \lambda)\propto \lambda^{0.5K}\exp(-0.5\lambda\btheta^{\top}P\btheta)$. In the notation $p(\lambda \vert \bfomega_{-j}, \mathcal{D})$, it is understood that $\lambda$ plays the role of the $j$th component $\omega_j$ of $\bfomega$ and thus $\bfomega_{-j}$ represents all the remaining model parameters except $\lambda$. Writing the full expression, we recover $p(\lambda \vert \bfomega_{-j}, \mathcal{D})\propto \lambda^{0.5(K+\nu)-1}\exp(-\lambda 0.5 (\btheta^{\top}P\btheta+\nu \delta))$, which is the kernel of a Gamma density $(\lambda \vert \bfomega_{-j}, \mathcal{D}) \sim \mathcal{G}(0.5(K+\nu),0.5 (\btheta^{\top}P\btheta+\nu \delta))$. For $\delta$, we have $p(\delta \vert \bfomega_{-j}, \mathcal{D})\propto p(\lambda \vert \delta) p(\delta)$, with $p(\lambda \vert \delta)\propto \delta^{0.5\nu}\exp(-0.5\lambda \nu \delta)$ and $p(\delta) \propto \delta^{a_{\delta}-1}\exp(-b_{\delta}\delta)$, yielding the conditional posterior $p(\delta \vert \bfomega_{-j}, \mathcal{D})\propto \delta^{0.5\nu+a_{\delta}-1} \exp(-\delta(0.5\lambda\nu+b_{\delta}))$. The latter is also the kernel of a Gamma density $(\delta \vert \bfomega_{-j}, \mathcal{D})\sim \mathcal{G}(0.5\nu + a_{\delta}, 0.5\lambda \nu + b_{\delta})$ and the Bayesian P-splines model is therefore partially conditionally conjugate.

\subsubsection{Conditional posteriors of spline coefficients}

\noindent The conditional posterior of the $k$th B-spline coefficient is  $p(\theta_k \vert \bfomega_{-j}, \mathcal{D}) \propto \mathcal{L}(\theta_k; \bfomega_{-j}, \mathcal{D})p(\theta_k \vert \btheta_{-k}, \lambda)$, where $\btheta_{-k}$ denotes vector $\btheta$ without the $k$th component. To explore the latter conditional posterior target with adaptive rejection sampling or Griddy-Gibbs sampling, it is preferable that the target can be evaluated in an efficient way. The complexity of evaluating the conditional likelihood $\mathcal{L}(\theta_k; \bfomega_{-j}, \mathcal{D})$ will directly depend on the model complexity, but the conditional prior $p(\theta_k \vert \btheta_{-k}, \lambda)$ is relatively straightforward to evaluate as it comes to evaluate a Gaussian density at $\theta_k$. This is because the joint (conditional) prior on $\btheta$ given $\lambda$ has univariate conditionals that are normally distributed. Proposition 1 provides closed-form expressions for the mean and variance of the conditional prior of $\theta_k$ for any spline component $k\in\{1,\dots,K\}$ and penalty order $r$. This result permits a simplified evaluation of the conditional prior, and, in turn, a simplified evaluation of $p(\theta_k \vert \bfomega_{-j}, \mathcal{D})$ seen as a function of $\theta_k$.\\

\noindent \textbf{Proposition 1.} For a penalty of order $r$, the conditional prior of the $k$th B-spline coefficient $\theta_k$ associated with the joint Gaussian prior $p(\btheta \vert \lambda) \propto \lambda^{K/2} \exp(-0.5\lambda \btheta^{\top}P\btheta)$ is normally distributed with mean $\mathbb{E}(\theta_k \vert \btheta_{-k},\lambda)=\psi_r(\btheta_{-k})z^{-1}_r(k,\varepsilon)$ and variance $\mathbb{V}(\theta_k \vert \btheta_{-k},\lambda)=(\lambda z_r(k,\varepsilon))^{-1}$, where $\psi_r(\btheta_{-k})$ depends on  $\btheta_{-k}$ and $z_r(k,\varepsilon)$ is the $k$th component of the diagonal of $P$.\\

\noindent \textbf{Proof.} See Appendix A1.\\

\noindent In particular, Proposition 1 can be used to write the analytical formulas of the conditional prior means and variances of the B-spline coefficients for $r=2$ and $r=3$. The associated terms $\psi_r(\btheta_{-k})$ and $z_r(k,\varepsilon)$ are given below for the sake of illustration.

\begin{eqnarray}
	\psi_r(\btheta_{-k})=
	\begin{cases}
		(2\theta_{k+1}-\theta_{k+2})\mathbb{I}(k=1)+(4\theta_{k+1}+2\theta_{k-1}-\theta_{k+2})\mathbb{I}(k=2) \nonumber \\
		+\big(4(\theta_{k-1}+\theta_{k+1})-(\theta_{k-2}+\theta_{k+2})\big)\mathbb{I}(3 \leq k \leq K-2) \nonumber \\
		+(4\theta_{k-1}+2\theta_{k+1}-\theta_{k-2}) \mathbb{I}(k=K-1)+(2\theta_{k-1}-\theta_{k-2})\mathbb{I}(k=K)\ \text{if}\ r=2, \nonumber \\
		\\
		(3(\theta_{k+1}-\theta_{k+2})+\theta_{k+3})\mathbb{I}(k=1)+(12\theta_{k+1}-6\theta_{k+2}+3\theta_{k-1}+\theta_{k+3})\mathbb{I}(k=2) \nonumber \\
		+(15\theta_{k+1}+12\theta_{k-1}-6\theta_{k+2}-3\theta_{k-2}+\theta_{k+3})\mathbb{I}(k=3) \nonumber \\
		+\big(15(\theta_{k-1}+\theta_{k+1})-6(\theta_{k-2}+\theta_{k+2})+(\theta_{k-3}+\theta_{k+3})\big)\mathbb{I}(4 \leq k \leq K-3) \nonumber \\
		+(15\theta_{k-1}+12\theta_{k+1}-6\theta_{k-2}-3\theta_{k+2}+\theta_{k-3})\mathbb{I}(k=K-2) \nonumber \\ +(12\theta_{k-1}-6\theta_{k-2}+3\theta_{k+1}+\theta_{k-3})\mathbb{I}(k=K-1)+(3(\theta_{k-1}-\theta_{k-2})+\theta_{k-3})\mathbb{I}(k=K)\ \text{if}\ r=3. \nonumber
	\end{cases}
\end{eqnarray}

\begin{eqnarray}
	z_r(k,\varepsilon)=
	\begin{cases}
		(1+\varepsilon)\mathbb{I}(k\in\{1,K\})+(5+\varepsilon)\mathbb{I}(k\in\{2,K-1\})+(6+\varepsilon)\mathbb{I}(3\leq k \leq K-2)\ \text{if}\ r=2, \\
		\\
		(1+\varepsilon)\mathbb{I}(k\in\{1,K\})+(10+\varepsilon)\mathbb{I}(k\in\{2,K-1\})+(19+\varepsilon)\mathbb{I}(k\in\{3,K-2\}) \nonumber \\
		+(20+\varepsilon)\mathbb{I}(4\leq k \leq K-3)\ \text{if}\ r=3. \nonumber  
	\end{cases}
\end{eqnarray}

\vspace{0.2cm}

\noindent From an algorithmic perspective, it is not really convenient to work with the indicator functions $\mathbb{I}(\cdot)$ to evaluate the conditional prior mean and variance of the $k$th B-spline coefficient. For a second-order penalty, define the diagonal $K\times K$ matrix $E_2=\text{diag}((1+\varepsilon)^{-1},(5+\varepsilon)^{-1},(6+\varepsilon)^{-1},\dots,(6+\varepsilon)^{-1},(5+\varepsilon)^{-1},(1+\varepsilon)^{-1})$, where the term $(6+\varepsilon)^{-1}$ is repeated in total $K-4$ times. For a third-order penalty, define the diagonal $K\times K$ matrix $E_3=\text{diag}((1+\varepsilon)^{-1},(10+\varepsilon)^{-1},(19+\varepsilon)^{-1},(20+\varepsilon)^{-1},\dots,(20+\varepsilon)^{-1},(19+\varepsilon)^{-1},(10+\varepsilon)^{-1},(1+\varepsilon)^{-1})$, where the term $(20+\varepsilon)^{-1}$ is repeated in total $K-6$ times. Moreover, define the $K\times K$ matrices $A_r=-D_r^{\top}D_r$ for $r\in\{2,3\}$ and $C_r=(A_r- A_r \odot I_K)E_r$, where $\odot$ denotes the Hadamard product and $I_K$ is the $K$-dimensional identity matrix. Then, the conditional prior mean of $\theta_k$ for penalty order $r$ can be obtained via the vector product $\mathbb{E}(\theta_k \vert \btheta_{-k},\lambda)=C_r^{\top}(k,\cdot)\btheta$, where $C_r^{\top}(k,\cdot)$ denotes the $k$th row of matrix $C_r^{\top}$. The variance is obtained as $\mathbb{V}(\theta_k \vert \btheta_{-k},\lambda)=E_r(k,k)\lambda^{-1}$, where $E_r(k,k)$ is the entry in $k$th row and $k$th column of matrix $E_r$.\\
\indent Denote the conditional log-likelihood function of the $k$th B-spline coefficient as $\ell(\theta_k; \bfomega_{-j}, \mathcal{D}):=\log \mathcal{L}(\theta_k; \bfomega_{-j}, \mathcal{D})$. Working with the closed-form expressions of Proposition 1, it is straightforward to show that the conditional posterior of $\theta_k$ is strictly log-concave whenever the following inequality holds $\partial^2\ell(\theta_k; \bfomega_{-j}, \mathcal{D})/\partial \theta_k^2 \leq 0$. This result is given in Proposition 2.\\

\noindent \textbf{Proposition 2.} If $\partial^2\ell(\theta_k; \bfomega_{-j}, \mathcal{D})/\partial \theta_k^2 \leq 0$, then $p(\theta_k \vert \bfomega_{-j}, \mathcal{D})$ is strictly log-concave.\\

\noindent \textbf{Proof.} We use basic properties of real-valued functions to show that the log conditional posterior distribution of the $k$th B-spline coefficient $\varphi_k(\theta_k):= \log p(\theta_k \vert \bfomega_{-j}, \mathcal{D}): \mathbb{R} \to \mathbb{R}$ is strictly concave. The function of interest is denoted by:

\vspace{-0.4cm}

\begin{eqnarray}
	\varphi_k(\theta_k)&\dot{=}&\log p_G\Big(\theta_k;\psi_r(\btheta_{-k})(z_r(k, \varepsilon))^{-1},(\lambda z_r(k, \varepsilon))^{-1}\Big)+\ell(\theta_k; \bfomega_{-j}, \mathcal{D}) \nonumber \\
	&\dot{=}&-\frac{\lambda}{2}z_r(k, \varepsilon)\Bigg(\theta_k-\frac{\psi_r(\btheta_{-k})}{z_r(k, \varepsilon)}\Bigg)^2+\ell(\theta_k; \bfomega_{-j}, \mathcal{D}) \nonumber \\
	&\dot{=}&-\frac{\lambda}{2}z_r(k, \varepsilon)\theta_k^2+\lambda \psi_r(\btheta_{-k})\theta_k+\ell(\theta_k; \bfomega_{-j}, \mathcal{D}), \nonumber 
\end{eqnarray}

\noindent where $\dot{=}$ denotes equality up to an additive constant. Since $\ell(\theta_k; \bfomega_{-j}, \mathcal{D})$ is assumed to be twice differentiable with respect to the spline coefficients, the following derivatives exist:

\vspace{-0.3cm}

\begin{eqnarray}
	\partial \varphi_k(\theta_k)/\partial \theta_k &=& \varphi_k'(\theta_k)=-\lambda z_r(k, \varepsilon) \theta_k+\lambda \psi_r(\btheta_{-k})+ \partial \ell(\theta_k; \bfomega_{-j}, \mathcal{D})/\partial \theta_k, \nonumber \\
	\nonumber \\
	\partial^2 \varphi_k(\theta_k)/\partial \theta_k^2 &=&\varphi_k''(\theta_k)=-\lambda z_r(k, \varepsilon)+\partial^2 \ell(\theta_k; \bfomega_{-j}, \mathcal{D})/\partial \theta_k^2. \nonumber
\end{eqnarray}

\noindent Note that $\lambda z_r(k, \varepsilon)>0$ since $\lambda>0$ and $\varepsilon>0$. Also, $\partial^2\ell(\theta_k; \bfomega_{-j}, \mathcal{D})/\partial \theta_k^2\leq 0$ by assumption. It follows that $\varphi_k''(\theta_k)<0$, a sufficient condition for strict concavity of $\varphi_k$. Said differently, $p(\theta_k \vert \bfomega_{-j}, \mathcal{D})$ is a strictly log-concave function of $\theta_k$ and therefore admits a unique maximum denoted by $\theta_k^*=\text{argmax}_{\theta_k \in \mathbb{R}} \varphi_k(\theta_k)$ $\square$\\

\noindent It may be convenient to compute $\theta_k^*=\text{argmax}_{\theta_k \in \mathbb{R}} \varphi_k(\theta_k)$ as a departure point to explore $p(\theta_k \vert \bfomega_{-j}, \mathcal{D})$ in a neighborhood of $\theta_k^*$. Instead of searching for the maximum on the entire real line $(-\infty, +\infty)$, it is possible to narrow down the search to a bounded subset of $\mathbb{R}$ as shown in Proposition 3. This result is particularly useful if the maximization problem is seen as a root-finding problem.\\

\noindent \textbf{Proposition 3.} Assume that $\partial^2\ell(\theta_k; \bfomega_{-j}, \mathcal{D})/\partial \theta_k^2 \leq 0$, so that the log conditional posterior $\varphi_k(\theta_k)$ is strictly concave. Then, for any $\kappa >0$, $\varphi'_k(0)<0$ is a sufficient condition for the conditional posterior mode $\theta^*_k$ to be located in the negative half-open interval $\zeta_{-}=\big[(\lambda z_r(k,\varepsilon))^{-1} \varphi'_k(0)-\kappa, 0\big[$ and $\varphi'_k(0)>0$ is a sufficient condition for the modal value to be located in the positive half-open interval $\zeta_{+}=\big]0, (\lambda z_r(k,\varepsilon))^{-1} \varphi'_k(0)+\kappa \big]$.\\

\noindent \textbf{Proof.} Let $\ell'(\theta_k; \bfomega_{-j}, \mathcal{D}) = \partial \ell(\theta_k; \bfomega_{-j}, \mathcal{D})/\partial \theta_k$. The first derivative of $\varphi_k(\theta_k)$ is given by:

\vspace{-0.3cm}

\begin{eqnarray}
	\varphi_k'(\theta_k)=-\lambda z_r(k, \varepsilon) \theta_k+\lambda \psi_r(\btheta_{-k})+\ell'(\theta_k; \bfomega_{-j}, \mathcal{D}), \nonumber
\end{eqnarray}

\vspace{0.3cm}

\noindent and its evaluation at $\theta_k=0$ is:

\vspace{-0.4cm}

\begin{eqnarray}
	\varphi_k'(0)= \lambda \psi_r(\btheta_{-k})+\ell'(\theta_k; \bfomega_{-j}, \mathcal{D})\Big\vert_{\theta_k=0}. \nonumber 
\end{eqnarray}

\vspace{0.2cm}

\noindent We start by considering the case $\varphi'_k(0)<0$. Since $\varphi_k$ is a strictly concave function in $\mathbb{R}$, the first derivative $\varphi'_k$ is a strictly decreasing function in $(-\infty, +\infty)$. Thus, when $\varphi'_k(0)<0$, the critical point $\theta_k^*$ (satisfying $\varphi'_k(\theta_k^*)=0$) must be located in $]-\infty, 0[$. Finding a lower bound for $\theta_k^*$ is equivalent to find a $\tilde{\theta}_k \in ]-\infty,0[$ such that:

\vspace{-0.3cm}

\begin{eqnarray}
	\varphi_k'(\tilde{\theta}_k)=-\lambda z_r(k, \varepsilon) \tilde{\theta}_k+\lambda \psi_r(\btheta_{-k})+\ell'(\theta_k; \bfomega_{-j}, \mathcal{D})\Big\vert_{\theta_k=\tilde{\theta}_k} > 0 \nonumber \\
	\Leftrightarrow \tilde{\theta}_k < (\lambda z_r(k,\varepsilon))^{-1}\left(\lambda \psi_r(\btheta_{-k})+\ell'(\theta_k; \bfomega_{-j}, \mathcal{D})\Big\vert_{\theta_k=\tilde{\theta}_k}\right). \nonumber 
\end{eqnarray}

\vspace{0.1cm}

\noindent By assumption, the conditional log-likelihood function is concave, i.e.\ $\partial^2\ell(\theta_k; \bfomega_{-j}, \mathcal{D})/\partial \theta_k^2 \leq 0$. Hence, the first derivative $\partial \ell(\theta_k; \bfomega_{-j}, \mathcal{D})/\partial \theta_k$ is a decreasing function in $\mathbb{R}$ and the following inequality holds for any $\tilde{\theta}_k<0$:

\vspace{-0.2cm}

\begin{eqnarray}
	\ell'(\theta_k; \bfomega_{-j}, \mathcal{D})\Big\vert_{\theta_k=0}\leq\ell'(\theta_k; \bfomega_{-j}, \mathcal{D})\Big\vert_{\theta_k=\tilde{\theta}_k}. \nonumber 
\end{eqnarray}

\vspace{0.2cm}

\noindent This implies that we also have the following inequality:

\vspace{-0.2cm}

\begin{eqnarray}
	&&(\lambda z_r(k,\varepsilon))^{-1}\left(\lambda \psi_r(\btheta_{-k})+\ell'(\theta_k; \bfomega_{-j}, \mathcal{D})\Big\vert_{\theta_k=0}\right) \leq (\lambda z_r(k,\varepsilon))^{-1}\left(\lambda \psi_r(\btheta_{-k})+\ell'(\theta_k; \bfomega_{-j}, \mathcal{D})\Big\vert_{\theta_k=\tilde{\theta}_k} \right). \nonumber 
\end{eqnarray}

\vspace{0.2cm}

\noindent Note that the term on the left-hand side is equal to $(\lambda z_r(k,\varepsilon))^{-1} \varphi_k'(0)$ and the inequality becomes:

\vspace{-0.2cm}

\begin{eqnarray}
	(\lambda z_r(k,\varepsilon))^{-1} \varphi_k'(0) \leq
	(\lambda z_r(k,\varepsilon))^{-1}\left(\lambda \psi_r(\btheta_{-k})+\ell'(\theta_k; \bfomega_{-j}, \mathcal{D})\Big\vert_{\theta_k=\tilde{\theta}_k}\right). \nonumber 
\end{eqnarray}

\noindent Therefore, choosing a $\tilde{\theta}_k$ satisfying $\tilde{\theta}_k<(\lambda z_r(k,\varepsilon))^{-1} \varphi_k'(0)$ implies that:

\vspace{-0.2cm}

\begin{eqnarray}
	\tilde{\theta}_k < (\lambda z_r(k,\varepsilon))^{-1}\left(\lambda \psi_r(\btheta_{-k})+\ell'(\theta_k; \bfomega_{-j}, \mathcal{D})\Big\vert_{\theta_k=\tilde{\theta}_k}\right), \nonumber 
\end{eqnarray}

\noindent which in turn implies $\varphi'_k(\tilde{\theta}_k)>0$. One particular such point is $\tilde{\theta}_k=(\lambda z_r(k,\varepsilon))^{-1} \varphi_k'(0)-\kappa$ for any finite $\kappa>0$. As such, the critical point belongs to the (strictly) negative half-open interval $\zeta_{-}=\big[(\lambda z_r(k,\varepsilon))^{-1} \varphi'_k(0)-\kappa, 0\big[$.\\

\noindent We now consider the case $\varphi'_k(0)>0$. Since $\varphi_k$ is a strictly concave function in $\mathbb{R}$, the first derivative $\varphi'_k$ is a strictly decreasing function in $(-\infty, +\infty)$. Thus, when $\varphi'_k(0)>0$, the critical point $\theta_k^*$ (satisfying $\varphi'_k(\theta_k^*)=0$) must be located in $]0,+\infty[$. Finding an upper bound for $\theta_k^*$ is equivalent to find a $\tilde{\theta}_k \in ]0,+\infty[$ such that:

\vspace{-0.25cm}

\begin{eqnarray}
	\varphi_k'(\tilde{\theta}_k)=-\lambda z_r(k, \varepsilon) \tilde{\theta}_k+\lambda \psi_r(\btheta_{-k})+\ell'(\theta_k; \bfomega_{-j}, \mathcal{D})\Big\vert_{\theta_k=\tilde{\theta}_k} < 0 \nonumber \\
	\Leftrightarrow \tilde{\theta}_k > (\lambda z_r(k,\varepsilon))^{-1}\left(\lambda \psi_r(\btheta_{-k})+\ell'(\theta_k; \bfomega_{-j}, \mathcal{D})\Big\vert_{\theta_k=\tilde{\theta}_k}\right). \nonumber 
\end{eqnarray}

\vspace{0.2cm}

\noindent By assumption, we have $\partial^2\ell(\theta_k; \bfomega_{-j}, \mathcal{D})/\partial \theta_k^2 \leq 0$. Hence, the first derivative $\partial \ell(\theta_k; \bfomega_{-j}, \mathcal{D})/\partial \theta_k$ is a decreasing function in $\mathbb{R}$ and the following inequality holds for any $\tilde{\theta}_k>0$:

\vspace{-0.2cm}

\begin{eqnarray}
	\ell'(\theta_k; \bfomega_{-j}, \mathcal{D})\Big\vert_{\theta_k=0}\geq\ell'(\theta_k; \bfomega_{-j}, \mathcal{D})\Big\vert_{\theta_k=\tilde{\theta}_k}. \nonumber 
\end{eqnarray}

\vspace{0.2cm}

\noindent This implies that we also have the following inequality:

\vspace{-0.3cm}

\begin{eqnarray}
	&&(\lambda z_r(k,\varepsilon))^{-1}\left(\lambda \psi_r(\btheta_{-k})+\ell'(\theta_k; \bfomega_{-j}, \mathcal{D})\Big\vert_{\theta_k=0}\right) \geq (\lambda z_r(k,\varepsilon))^{-1}\left(\lambda \psi_r(\btheta_{-k})+\ell'(\theta_k; \bfomega_{-j}, \mathcal{D})\Big\vert_{\theta_k=\tilde{\theta}_k} \right). \nonumber 
\end{eqnarray}

\vspace{0.2cm}

\noindent Note that the term on the left-hand side is equal to $(\lambda z_r(k,\varepsilon))^{-1} \varphi_k'(0)$ and the inequality becomes:

\vspace{-0.2cm}

\begin{eqnarray}
	(\lambda z_r(k,\varepsilon))^{-1} \varphi_k'(0) \geq
	(\lambda z_r(k,\varepsilon))^{-1}\left(\lambda \psi_r(\btheta_{-k})+\ell'(\theta_k; \bfomega_{-j}, \mathcal{D})\Big\vert_{\theta_k=\tilde{\theta}_k}\right). \nonumber 
\end{eqnarray}

\vspace{0.15cm}

\noindent Therefore, choosing a $\tilde{\theta}_k$ satisfying $\tilde{\theta}_k>(\lambda z_r(k,\varepsilon))^{-1} \varphi_k'(0)$ implies that:

\vspace{-0.1cm}

\begin{eqnarray}
	\tilde{\theta}_k > (\lambda z_r(k,\varepsilon))^{-1}\left(\lambda \psi_r(\btheta_{-k})+\ell'(\theta_k; \bfomega_{-j}, \mathcal{D})\Big\vert_{\theta_k=\tilde{\theta}_k}\right), \nonumber 
\end{eqnarray}

\noindent which in turn implies $\varphi'_k(\tilde{\theta}_k)<0$. One particular such point is $\tilde{\theta}_k=(\lambda z_r(k,\varepsilon))^{-1} \varphi_k'(0)+\kappa$ for any finite $\kappa>0$. As such, the critical point belongs to the (strictly) positive half-open interval $\zeta_{+}=\big]0, (\lambda z_r(k,\varepsilon))^{-1} \varphi'_k(0)+\kappa\big]\ \square$

\subsubsection{Conditional posteriors of other model coefficients}

\noindent The conditional posteriors of additional hyperparameters in $\bfeta$ and unpenalized parameters in $\bgamma$ are treated in a similar way as the B-spline coefficients in Section 2.3.2. The classic Gaussian prior assumption on unpenalized parameters $\bgamma \sim \mathcal{N}(\mu_{\bgamma},\Sigma_{\bgamma})$ ensures that the conditional priors are normally distributed and therefore evaluation of the conditional posterior $p(\gamma_k \vert \bfomega_{-j}, \mathcal{D})$, $k\in\{1,\dots,l\}$ boils down to evaluating the conditional likelihood $\mathcal{L}(\gamma_k; \bfomega_{-j}, \mathcal{D})$ and a Gaussian conditional prior. Note that under an uninformative prior with $\mu_{\bgamma}=0$ and a diagonal covariance matrix $\Sigma_{\bgamma}$ with diagonal entries equal to $\tau^{-1}$, where $\tau>0$ is a small precision parameter, the conditional prior is $(\gamma_k \vert \bgamma_{-k})\sim \mathcal{N}(0,\tau^{-1})$ and its evaluation is trivial. Furthermore, under a Gaussian prior for $\bgamma$, we can also use Proposition 2 which ensures that $p(\gamma_k \vert \bfomega_{-j}, \mathcal{D})$ will be strictly log-concave provided $\partial^2\ell(\gamma_k; \bfomega_{-j}, \mathcal{D})/\partial \gamma_k^2\leq 0$, a condition that is usually easy to check for most models encountered in practice. If the Bayesian P-splines model contains hyperparameters (other than $\lambda$ and $\delta$) that are not conditionally conjugate and also non-Gaussian, it may become difficult (if not impossible) to check whether or not the associated conditionals are log-concave. In that case, the adaptive rejection sampler of \cite{gilks1992adaptive} is not an option anymore and we advocate the Griddy-Gibbs sampler of \cite{ritter1992facilitating} as a fallback option. The next section gives a detailed exposition of the Gibbs sampler for Bayesian P-splines (GSBPS) along with a general pseudo-code that can be used to obtain a Markov chain sample from the joint posterior $p(\bfomega \vert \mathcal{D})$.

\section{Gibbs sampler for Bayesian P-splines (GSBPS)}

\noindent We use the properties of the univariate conditionals discussed in Section 2 to build a simple and appealing Gibbs sampler. The proposed Gibbs sampler for Bayesian P-splines (GSBPS) permits to obtain a sample from the joint posterior $p(\bfomega \vert \mathcal{D})$ by cycling through the (univariate) conditional posteriors. When it can be shown that the conditional posterior $p(\omega_j \vert \bfomega_{-j}, \mathcal{D})$ is log-concave, we recommend using the black box technique of \cite{gilks1992adaptive}; i.e.\ the adaptive rejection sampler; which is an efficient technique to sample from a log-concave posterior target. When $p(\omega_j \vert \bfomega_{-j}, \mathcal{D})$ is not log-concave or when it is difficult to mathematically show log concavity, we suggest to rely on the Griddy-Gibbs sampler of \cite{ritter1992facilitating}, which is computationally more demanding but permits to sample a posterior target even if it is not log-concave. Denote by $L_{\mathcal{C}}$ the set of log-concave functions. We use the set-theoretic notation $\omega_j \in L_{\mathcal{C}}$ to indicate that the model parameter $\omega_j$ has a conditional posterior $p(\omega_j \vert \bfomega_{-j}, \mathcal{D})$ that is log-concave and $\omega_j \notin L_{\mathcal{C}}$ otherwise. When $\omega_j$ is either a B-spline coefficient in $\btheta$ or a non-penalized parameter in $\bgamma$, it has been shown previously that $\omega_j \in L_{\mathcal{C}}$ whenever $\partial \log \ell(\omega_j; \bfomega_{-j}; \mathcal{D})/\partial \omega_j \leq 0$ holds (i.e.\ the conditional log-likelihood function is concave). When this is the case, the corresponding univariate conditional is sampled by means of the adaptive rejection sampling algorithm described below.

\subsection{Adaptive rejection sampling}

\noindent We do not attempt to give a detailed explanation of the adaptive rejection sampling algorithm of \cite{gilks1992adaptive}. Instead, a short exposition is provided, focusing on the main ingredients involved that are required to implement the sampler. The reader is redirected to \cite{gilks1992adaptive} for technical details and to \cite{wild1993algorithm} for a full treatment of the algorithm.\\
\indent Assume that $p(\omega_j \vert \bfomega_{-j}, \mathcal{D})$ is log-concave ($\omega_j \in L_{\mathcal{C}}$) and denote by $\varphi_j(\omega_j)= \log p(\omega_j \vert \bfomega_{-j}, \mathcal{D})$ the log conditional posterior target seen as a function of $\omega_j$ in $\mathbb{R}$. The algorithm requires to initialize a set of $L\in \mathbb{N}$ abscissae points in $\mathbb{R}$, denoted by $\mathcal{A}_L=\{\omega_{j(1)},\dots,\omega_{j(L)}\}$, with $\omega_{j(1)}\leq \omega_{j(2)} \leq \dots \leq \omega_{j(L)}$. \cite{gilks1992adaptive} have shown that $L=2$ is necessary and sufficient for computational efficiency, but nothing restrains to choose a larger number. Furthermore, since $\omega_j$ evolves in an unbounded domain, there must be starting points on both sides of the mode of $p(\omega_j \vert \bfomega_{-j}, \mathcal{D})$ \citep{wild1993algorithm}. To initialize $\mathcal{A}_L$, we compute $\omega_j^*=\text{argmax}_{\omega_j \in \mathbb{R}}\varphi_j(\omega_j)$ (uniqueness is guaranteed under strict concavity) by using, for instance, a root-finding algorithm or a Newton-Raphson algorithm. Then, we do a Laplace approximation to $p(\omega_j \vert \bfomega_{-j}, \mathcal{D})$ centered around $\omega_j^*$, i.e.\ $\widetilde{p}_G(\omega_j \vert \bfomega_{-j}, \mathcal{D})=\mathcal{N}(\omega_j^*, \sigma_j^{*2})$, with $\sigma_j^*=\sqrt{(-\varphi_j''(\omega_j^*))^{-1}}$. Next, we compute the endpoints in $\mathcal{A}_L$ as $\omega_{j(1)}=\omega_j^*-c\sigma_j^*$ and $\omega_{j(L)}=\omega_j^*+c\sigma_j^*$ for some constant $c>0$ (e.g.\ $c=2$) and construct $\mathcal{A}_L$ by taking $L=5$ equidistant points between (and including) the endpoints. The adaptive rejection sampler requires to define two functions based on $\varphi_j(\omega_j)$, namely a lower hull and an upper hull. For $\omega_j \in [\omega_{j(1)},\omega_{j(L)}[$, the lower hull corresponds to the following piecewise linear function with breakpoints at abscissae points in $\mathcal{A}_L$:   

\vspace{-0.2cm}

\begin{eqnarray}
h_L^{\text{low}}(\omega_j)&=&\sum_{l=1}^{L-1} \Bigg(\frac{\varphi_j(\omega_{j(l)})(\omega_{j(l+1)}-\omega_j) + \varphi_j(\omega_{j(l+1)})(\omega_j-\omega_{j(l)})}{\omega_{j(l+1)}-\omega_{j(l)}} \mathbb{I}(\omega_{j(l)}\leq \omega_j < \omega_{j(l+1)})\Bigg). \nonumber 
\end{eqnarray}

\noindent For $\omega_j = \omega_{j(L)}$, we have $h_L^{\text{low}}(\omega_j)=\varphi(\omega_{j(L)})$ and $h_L^{\text{low}}(\omega_j)=-\infty$ for $\omega_j \notin [\omega_{j(1)},\omega_{j(L)}]$. The upper hull is a piecewise linear function with breakpoints corresponding to intersection points of the tangents to $\varphi(\cdot)$ at the abscissae in $\mathcal{A}_L$. It can be shown that the $L-1$ breakpoints of the upper hull are given by:

\vspace{-0.2cm}

\begin{eqnarray}
z_l=\frac{\varphi_j(\omega_{j(l+1)})-\varphi_j(\omega_{j(l)})+\omega_{j(l)}\varphi'_j(\omega_{j(l)})-\omega_{j(l+1)}\varphi'_j(\omega_{j(l+1)})}{\varphi'_j(\omega_{j(l)})-\varphi'_j(\omega_{j(l+1)})},\ l=1,\dots,L-1. \nonumber 
\end{eqnarray}

\noindent The above breakpoints can be used to define the upper hull:

\vspace{-0.2cm}

\begin{eqnarray}
h_L^{\text{up}}(\omega_j)&=&\Big(\varphi_j(\omega_{j(1)}) + (\omega_j-\omega_{j(1)})\varphi'_j(\omega_{j(1)})\Big) \mathbb{I}(-\infty < \omega_j < z_1) \nonumber \\
&&+\sum_{l=1}^{L-2}\Bigg(\varphi_j(\omega_{j(l+1)}) + (\omega_j-\omega_{j(l+1)})\varphi'_j(\omega_{j(l+1)})\Bigg) \mathbb{I}(z_l \leq \omega_j < z_{l+1}) \nonumber \\
&&+ \Big(\varphi_j(\omega_{j(L)}) + (\omega_j-\omega_{j(L)})\varphi'_j(\omega_{j(L)})\Big) \mathbb{I}(z_{L-1} \leq \omega_j < +\infty). \nonumber
\end{eqnarray}

\noindent From there, compute the normalizing factor of the exponentiated upper hull $c_{\text{up}}=\int_{-\infty}^{+\infty}h_L^{\text{up}}(s)ds$ and define $g_L(\omega_j)=c_{\text{up}}^{-1}\exp(h_L^{\text{up}}(\omega_j))$. A sample $\omega_j$ from the conditional posterior $p(\omega_j \vert \bfomega_{-j}, \mathcal{D})$ can be obtained by following a two-step procedure. (Step 1: Sampling) In the sampling step, start by independently drawing a candidate $\omega_j$ from $g_L(\cdot)$ and a value $u$ from a uniform distribution in $(0,1)$. If $u\leq \exp(h_L^{\text{low}}(\omega_j)-h_L^{\text{up}}(\omega_j))$, the candidate $\omega_j$ is accepted, otherwise $\varphi_j(\omega_j)$ is evaluated and the candidate $\omega_j$ is accepted provided $u\leq \exp(\varphi_j(\omega_j)-h_L^{\text{up}}(\omega_j))$. When the latter inequality is not satisfied, $\omega_j$ is rejected and the algorithm proceeds to the updating step. (Step 2: Updating) Include the generated candidate $\omega_j$ in $\mathcal{A}_L$ and denote by $\mathcal{A}_{L+1}$ the new set of abscissae points with elements arranged in ascending order. Finally, update the lower and upper hull functions based on $\mathcal{A}_{L+1}$ and return to the sampling step. This two-step procedure is iterated until acceptance of a candidate.

\subsection{Griddy-Gibbs sampling}

\noindent When the conditional posterior $p(\omega_j \vert \bfomega_{-j}, \mathcal{D})$ is not log-concave ($\omega_j \notin L_{\mathcal{C}}$) or when the log concavity property is difficult to assess mathematically, we recommend the Griddy-Gibbs method of \cite{ritter1992facilitating}. The Griddy-Gibbs sampler provides a grid-based approximation of the distribution function of the target $p(\omega_j \vert \bfomega_{-j}, \mathcal{D})$ and the resulting approximation is used to generate a sample which can be seen as an approximate sample from the target. Again, we do not attempt to give a complete description of the algorithm since all the details can be found in the original Griddy-Gibbs paper.\\
\indent Central to Griddy-Gibbs is the set of $L\in \mathbb{N}$ abscissae points denoted by $\mathcal{B}_L=\{\omega_{j(1)},\dots,\omega_{j(L)}\}$, which will be used to approximate $p(\omega_j \vert \bfomega_{-j}, \mathcal{D})$. Elements in $\mathcal{B}_L$ are computed following a grid-grower method \citep{ritter1992facilitating} starting from $\omega_j^*=\text{argmax}_{\omega_j \in \mathbb{R}}\varphi_j(\omega_j)$ (note that a multimodal target $\varphi_j(\omega_j)$ is a possibility here). The grid-grower departs from $\omega_j^*$ and moves left (respectively) right until the following inequality holds $\varphi_j(\omega_j)-\varphi_j(\omega_j^*)<c_f$ for some contant $c_f>0$. In practice, setting $c_f=\log(0.01)$ usually ensures that the grid has grown far enough in the tails of the target and supports most of the conditional posterior probability mass. Setting $c_f=\log(0.01)$ means that the grid-grower proceeds by moving in the tails of the target until reaching $1\%$ of the maximum value $p(\omega_j^* \vert \bfomega_{-j}, \mathcal{D})$.\\ 
\indent We suggest using a grid-grower with growing steps of size $2^{k}\sqrt{(-\varphi_j''(\omega^*_j))^{-1}}$ with $k\in\{0,1,2,\dots\}$. As such, the step size accounts for the (local) curvature of the target around the modal value. Once the endpoints $\omega_{j(1)}$ and $\omega_{j(L)}$ have been obtained from the grid-grower, we construct $\mathcal{B}_L$ by taking $L=100$ equidistant points between (and including) the endpoints. Next, evaluate $\widetilde{\pi}_l=p(\omega_{j(l)} \vert \bfomega_{-j}, \mathcal{D})$ for $l=1,\dots,L$ and compute the probability masses $\pi_l=\widetilde{\pi}_l/\sum_{l=1}^L \widetilde{\pi}_l$ for $l=1,\dots,L$ representing the (discrete) approximation of the conditional posterior target distribution $p(\omega_j \vert \bfomega_{-j}, \mathcal{D})$. An approximate sample from the target is taken to be a candidate $\omega_{j(l)} \in \mathcal{B}_L$ drawn with probability $\pi_l$.

\newpage 

\subsection{The GSBPS algorithm}

\noindent The GSBPS algorithm proceeds as follows. Initial values are chosen for the model parameters and $\omega_j^{(0)}$ denotes the starting value of the $j$th component of the parameter vector $\bfomega$ of dimension $J$. Initial parameter values can be obtained by sampling the prior distributions \citep{bolstad2011understanding}. For $\btheta$ a possible initial value is the zero vector, although more suitable choices may exist depending on the model being considered (see Section 4). The prior mean of the penalty parameter is $\mathbb{E}(\lambda \vert \delta)=\delta^{-1}$, which motivates the initial value $\lambda^{(0)}=(\mathbb{E}(\delta))^{-1}=(a_{\delta}/b_{\delta})^{-1}$. A sample $\mathcal{S}_{\omega}=\{\bfomega^{(1)},\dots,\bfomega^{(M)}\}$ of size $M$ from $p(\bfomega \vert \mathcal{D})$ is obtained through a Gibbs strategy, cycling through the univariate conditionals as illustrated in the following pseudo-code.

\vspace{1cm}

\hrule
\vspace{0.1cm}
\noindent {\textbf{GSBPS algorithm to sample the joint posterior}\ $p(\bfomega \vert \mathcal{D})$} \label{GGS}
\vspace{0.1cm}
\hrule 
\vspace{0.1cm}
\noindent 1:\ Initialize model parameters\ $\omega_j^{(0)}$ and specify a chain length $M$.\\
2:\ \textbf{for} $m$ in 1 to $M$ \textbf{do}: \\
3:\ \hspace{0.4cm} Sample $\delta^{(m)}\sim\mathcal{G}(0.5\nu + a_{\delta},0.5\lambda^{(m-1)}\nu+b_{\delta})$.\\
4:\ \hspace{0.4cm} Sample $\lambda^{(m)}\sim\mathcal{G}(0.5(K+\nu), 0.5(\btheta^{(m-1)\top}P\btheta^{(m-1)}+\nu \delta^{(m)}))$.\\
5:\ \hspace{0.4cm} Replace $\delta$ and $\lambda$ components of $\bfomega^{(m-1)}$ by $\delta^{(m)}$ and $\lambda^{(m)}$, respectively.\\
6:\ \hspace{0.4cm} \textbf{for} $j$ in 1 to $J-2$ \textbf{do}: \\
7:\ \hspace{0.8cm} If $\omega_j \in L_{\mathcal{C}}$ (Adaptive rejection sampling)\\
8:\ \hspace{1.3cm} Construct the initial set of abscissae points $\mathcal{A}_L$.\\
9:\ \hspace{1.3cm} Sample from the normalized exponentiated upper hull $\omega_j^{(m)}\sim g_L(\omega_j)$.\\
10:\ \hspace{1.1cm} Sample $u \sim \mathcal{U}(0,1)$.\\
11:\ \hspace{1.1cm} If $u\leq \exp\left(h_L^{\text{low}}(\omega_j^{(m)})-h_L^{\text{up}}(\omega_j^{(m)})\right)$ accept $\omega_j^{(m)}$, else compute $\varphi_j(\omega_j^{(m)})$.\\
12:\ \hspace{1.1cm} If $u\leq \exp\left(\varphi_j(\omega_j^{(m)})-h_L^{\text{up}}(\omega_j^{(m)})\right)$ accept $\omega_j^{(m)}$. \\
13:\ \hspace{1.1cm} Else update lower and upper hulls by augmenting $\mathcal{A}_L$ with $\omega_j^{(m)}$ and return to step 9. \\
14:\ \hspace{1.1cm} Replace the $j$th component of $\bfomega^{(m-1)}$ by $\omega_j^{(m)}$. \\
15:\ \hspace{0.6cm} If $\omega_j \notin L_{\mathcal{C}}$ (Griddy-Gibbs sampling)\\
16:\ \hspace{1.1cm} Construct $\mathcal{B}_L=\{\omega_{j(1)},\dots,\omega_{j(L)}\}$ with the grid-grower method.\\
17:\ \hspace{1.1cm} Evaluate $\widetilde{\pi}_l=p(\omega_{j(l)} \vert \bfomega_{-j}^{(m-1)}, \mathcal{D})$ for $l=1,\dots,L$.\\
18:\ \hspace{1.1cm} Sample $\omega_j^{(m)}$ from $\mathcal{B}_L$ based on the probabilities $\pi_l=\widetilde{\pi}_l/\sum_{l=1}^L \widetilde{\pi}_l$,  $l=1,\dots,L$.\\
19:\ \hspace{1.1cm} Replace the $j$th component of $\bfomega^{(m-1)}$ by $\omega_j^{(m)}$. \\
20:\ \hspace{0.2cm} \textbf{end for}\\
21:\ \hspace{0.2cm} Set $\bfomega^{(m)}=\bfomega^{(m-1)}$. \\
22:\ \textbf{end for} \\
23:\ \textbf{Return}  $\mathcal{S}_{\omega}=\{\bfomega^{(1)},\dots,\bfomega^{(M)}\}$.
\vspace{0.1cm}
\hrule

\vspace{1cm}

\noindent The GSBPS algorithm starts by sampling the conditionally conjugate hyperparameters $\delta$ and $\lambda$. Then, depending on whether $\omega_j \in L_{\mathcal{C}}$ or not, the algorithm samples the remaining conditionals by alternating between the adaptive rejection sampler and the Griddy-Gibbs sampler. When the chain length $M$ is large enough and after having discarded early runs (burn-in), the Markov chain in $\mathcal{S}_{\omega}$ forms an approximate random sample from the joint posterior $p(\bfomega \vert \mathcal{D})$, which can be used to compute estimates of quantities involving the components in $\bfomega$. 

\newpage 

\section{Applications}

\subsection{Histogram smoothing and density estimation}

\noindent The Poisson distribution plays an important role in statistics. Its popularity to model discrete events is owed mostly to appealing mathematical properties and to the fact that a single parameter is required to characterize the distribution. Poisson assumptions on data are often encountered. This is for instance the case in density estimation by histogram smoothing, where counts associated to each histogram bin are assumed to be Poisson distributed. P-splines are particularly attractive for smoothing a series of counts \citep{eilers2021practical} and we illustrate how this can be achieved in a Bayesian context by means of the GSBPS algorithm.\\
\indent Consider a histogram with $n$ bins of equal width $\Delta>0$ and let $x_i\in \mathbb{R}$ denote the midpoint of the $i$th bin. The histogram bins are written as half-open intervals $\mathbb{B}_i=[x_i-\Delta/2, x_i+\Delta/2[$ for $i=1,\dots,n-1$ and the last bin is the closed interval $\mathbb{B}_n=[x_n-\Delta/2, x_n+\Delta/2]$. Let $y_i \in \mathbb{N} \cup \{0\}$ denote the count variable associated with the $i$th bin. Our model assumes that the observed counts are $n$ i.i.d.\ Poisson observations, each observation having unknown mean $\mu(x_i)>0$, denoted by $y_i \sim \mathcal{P}(\mu(x_i)),\ i=1,\dots,n$. B-splines are used to approximate the (unknown) function $\mu$. As $\mu$ is constrained to live on the (strictly) positive real line, a log-link function is used $\log \mu(x_i)=\sum_{k=1}^K \theta_k b_k(x_i)=\btheta^{\top}\boldsymbol{b}(x_i)$, where $\btheta = (\theta_1,\dots,\theta_K)^{\top} \in \mathbb{R}^K$ denotes the vector of B-spline coefficients to be estimated and $\boldsymbol{b}(\cdot)=(b_1(\cdot),\dots,b_K(\cdot))^{\top}$ is a cubic B-spline basis defined on a compact support $\mathcal{B}\supseteq\cup_{i=1}^n \mathbb{B}_i$. A default lower bound for $K$ is often taken to be ten \citep{eilers2021practical} and we therefore assume $K \geq 10$. Using our previous notation, the vector gathering all model parameters is $\bfomega=(\btheta^{\top}, \bfeta^{\top})^{\top}$ with $\bfeta^{\top}=(\lambda, \delta)$. The log-likelihood function based on the histogram values  $\mathcal{D}=(x_1,\dots,x_n, y_1,\dots,y_n)^{\top}$ is given by:

\vspace{-0.3cm}
	
\begin{eqnarray}
 \ell(\btheta; \mathcal{D}) \dot{=} \sum_{i=1}^n y_i \btheta^{\top}\boldsymbol{b}(x_i)-\sum_{i=1}^n \exp(\btheta^{\top}\boldsymbol{b}(x_i)). \nonumber 
\end{eqnarray}

\noindent It is easy to show that the conditional log-likelihood satisfies:

\vspace{-0.3cm}

\begin{eqnarray}
\frac{\partial^2 \ell(\theta_k; \btheta_{-k}, \mathcal{D})}{\partial \theta_k^2}=-\sum_{i=1}^n \exp(\theta_k b_k(x_i))\exp(\btheta_{-k}^{\top}\boldsymbol{b}_{-k}(x_i))b_k^2(x_i)\leq 0,\ \text{for}\ k=1,\dots,K, \nonumber 
\end{eqnarray}

\noindent where $\boldsymbol{b}_{-k}(\cdot)$ is the B-spline basis $\boldsymbol{b}(\cdot)$ omitting the $k$th component. From Proposition 2, it follows that the conditional posteriors $p(\theta_k \vert \btheta_{-k}, \lambda, \delta, \mathcal{D}),\ k=1,\dots,K$ are strictly log-concave and thus $\theta_k \in L_{\mathcal{C}}$ for all B-spline coefficients. In other words, the Griddy-Gibbs sampling step in the GSBPS algorithm is not required here to sample the joint posterior $p(\bfomega \vert \mathcal{D})$. The Markov chain for the B-spline coefficients $\btheta^{(1)},\dots,\btheta^{(M)}$ resulting from the GSBPS algorithm can be used to compute the point estimate $\widehat{\btheta}$ (based on the posterior mean) and the resulting smoothed histogram is $\widehat{\mu}(x)=\widehat{\btheta}^{\top}\boldsymbol{b}(x)$. The density estimate is obtained by computing the normalized smoothed histogram $\widehat{f}(x)=(\int_{-\infty}^{+\infty}\widehat{\mu}(s)ds)^{-1}\widehat{\mu}(x)$, where the integral is approximated numerically by means of a Riemann rule in the compact set $\mathcal{B}$. The GSBPS algorithm is implemented in the context of histogram smoothing by working with two datasets obtained from the \textit{JOPS} package in R. For the prior specification on the hyperparameters, we use $a_{\delta}=b_{\delta}=10^{-4}$ and $\nu=2$ (see Section 2.2) and generate a Markov chain of length $M=15 000$ from which we discard the first $5000$ iterations (burn-in) in the computation of the estimate $\widehat{\btheta}$. The penalty parameter is initialized at $\lambda^{(0)}=1$ and the initial value of $\btheta$ is taken to be $\btheta^{(0)}=(B^{\top}\text{diag}(y_1+1,\dots,y_n+1)B+\lambda^{(0)}P)^{-1}B^{\top}((y_1+1)\log(y_1+1),\dots,(y_n+1)\log(y_n+1))^{\top}$, where $B$ is the $n\times K$ basis matrix. For the datasets analyzed below, trace plots and Geweke diagnostic checks \citep{geweke1992evaluating} are available online (\url{https://github.com/oswaldogressani/GSBPS}). Globally, the trace plots show good mixing properties and Geweke diagnostics indicate convergence.

\subsubsection{Old Faithful geyser data}

\begin{figure}[h!]
	\begin{center}
		\includegraphics[width=16.5cm, height=6cm]{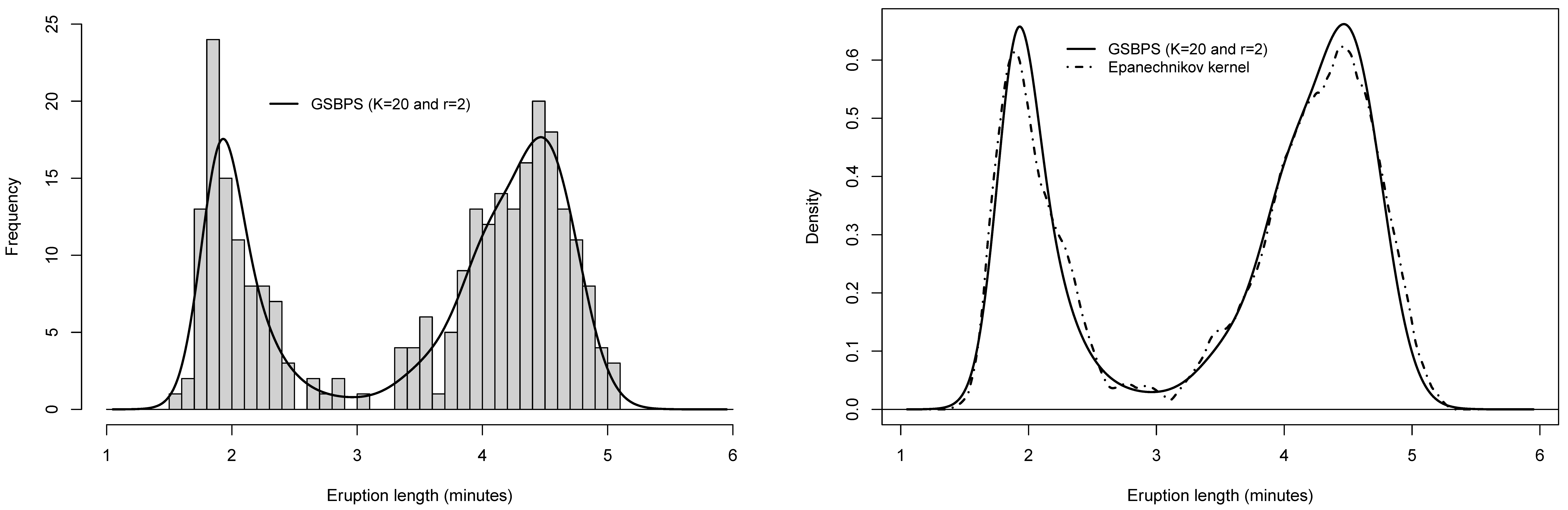}	
		\caption{Smoothed histogram (left panel) and density estimate (right panel) obtained with the GSBPS algorithm for the Old Faithful geyser data with $K=20$ and $r=2$. The dashed curve on the right panel is a density estimate obtained with an Epanechnikov kernel.}
		\label{fig1}
	\end{center}
\end{figure}

\noindent First, we consider a classic dataset on eruption times of the Old Faithful geyser in Yellowstone National Park that is used to construct a histogram with bin width $\Delta=0.1$. We use the Gibbs sampling scheme described in Section 3 to smooth the histogram with $K=20$ B-spline basis functions and a second-order penalty $r=2$. Figure~\ref{fig1} (left panel) shows the histogram of the Old Faithful data and its smooth version using the GSBPS algorithm. The estimated density shown in the right panel of Figure~\ref{fig1} is compared against a classic kernel estimate of the density based on an Epanechnikov kernel. The two density estimates exhibit similar patterns except for regions in the domain characterized by stronger curvature. Furthermore, we observe that the Bayesian P-splines model fits a slightly smoother curve as compared to the kernel method. 

\subsubsection{Hidalgo stamp data}

\begin{figure}[h!]
	\begin{center}
		\includegraphics[width=16.5cm, height=6cm]{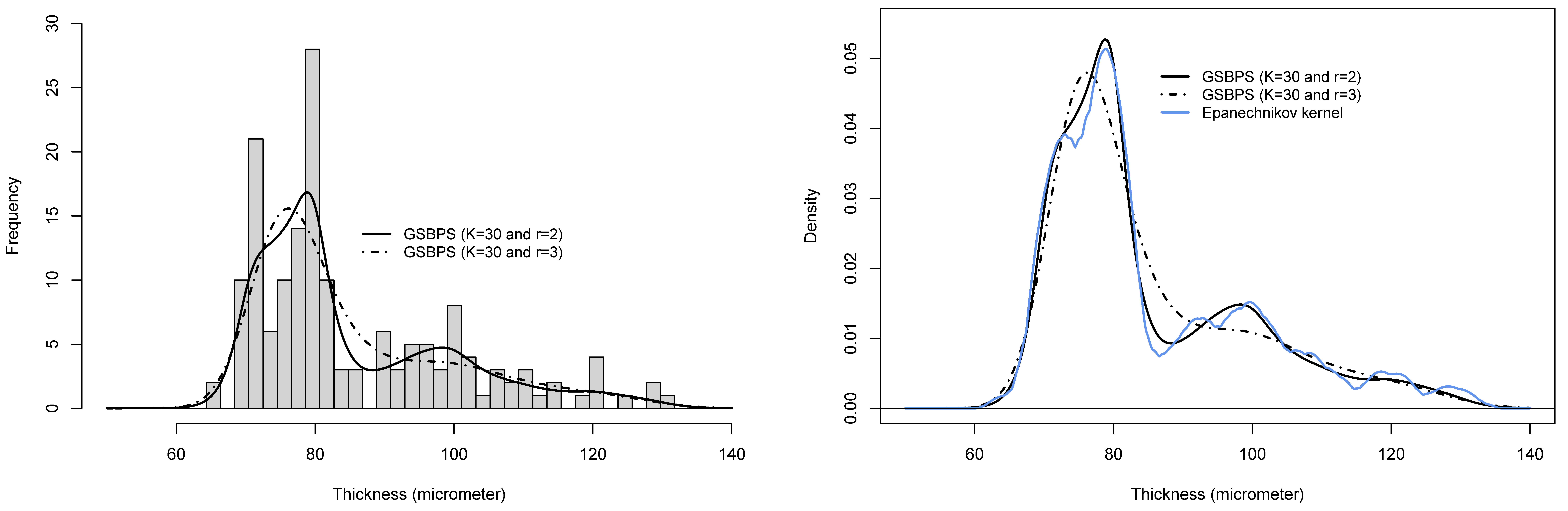}	
		\caption{Smoothed histogram (left panel) and density estimate (right panel) obtained with the GSBPS algorithm for the Hidalgo stamp data with $K=30$ and two penalty orders. The blue curve on the right panel is a density estimate obtained with an Epanechnikov kernel.}
		\label{fig2}
	\end{center}
\end{figure}

\noindent The second dataset is about thickness of stamp paper from the 1872 Hidalgo issue of Mexico \citep{basford1997modelling}. The histogram for this data is constructed with a bin width $\Delta=2.045$ obtained from the method of \cite{sheather1991reliable}. The Bayesian P-splines smoothed histogram is fitted with $K=30$ B-spline basis functions and with penalties of order $r=2$ and $r=3$, respectively, using the GSBPS algorithm. Results are shown in Figure~\ref{fig2}. The left panel highlights that the shape of the smoothed histogram varies with the penalty order and that a second-order penalty provides a more wiggly estimate. The kernel estimate of the density (obtained with an Epanechnikov kernel) is close to the fitted Bayesian P-splines density with $r=2$, although the latter is smoother.

\subsection{Binomial regression}

\noindent Consider a triplet $(y_i, m_i, x_i)$, where the discrete variable $y_i \in \{0,1,\dots,m_i\}$ represents the number of successes obtained in $m_i \in \mathbb{N}$ independent trials involving a covariate $x_i \in \mathbb{R}$. Also, assume that $\pi_i \in (0,1)$ denotes the probability of obtaining a success in each trial. In that case, it is well known that the random variable $Y_i$ has a binomial distribution $Y_i\sim\text{Bin}(m_i, \pi_i)$. 

\vspace{0.5cm}

\newpage 

\noindent Moreover, we use $\mathcal{D}=\big((y_1,m_1,x_1),\dots,(y_n,m_n,x_n)\big)^{\top}$ to denote a dataset resulting from the observation of $n$ triplets. Relying on a logit link, we express the relationship between $\pi_i$ and $x_i$ by using B-spline basis functions, i.e.\ $\log(\pi_i/(1-\pi_i))=\btheta^{\top}\boldsymbol{b}(x_i)$. As before, $\btheta = (\theta_1,\dots,\theta_K)^{\top} \in \mathbb{R}^K$ denotes the vector of B-spline coefficients and $\boldsymbol{b}(\cdot)=(b_1(\cdot),\dots,b_K(\cdot))^{\top}$ is a cubic B-spline basis defined on a compact support $\mathcal{B}\supseteq \mathcal{X}$, where $\mathcal{X}=[\min(x_1,\dots,x_n),\max(x_1,\dots,x_n)]$ represents the data range for the covariate. In a penalized setting, the vector gathering all model parameters is $\bfomega=(\btheta^{\top}, \bfeta^{\top})^{\top}$, with $\bfeta^{\top}=(\lambda, \delta)$. The log-likelihood function of this binomial model is:

\vspace{-0.3cm}

\begin{eqnarray}
\ell(\btheta; \mathcal{D})\dot{=}\sum_{i=1}^n\Big(y_i \btheta^{\top}\boldsymbol{b}(x_i)-m_i \log(1+\exp(\btheta^{\top}\boldsymbol{b}(x_i)))\Big). \nonumber
\end{eqnarray}

\noindent Note that the conditional log-likelihood satisfies:

\vspace{-0.3cm}

\begin{eqnarray}
	\frac{\partial^2 \ell(\theta_k; \btheta_{-k}, \mathcal{D})}{\partial \theta_k^2}=-\sum_{i=1}^n \left(\frac{m_i\exp(-\theta_k b_k(x_i))\exp(-\btheta_{-k}^{\top}\boldsymbol{b}_{-k}(x_i))}{\Big(1+\exp(-\theta_k b_k(x_i))\exp(-\btheta_{-k}^{\top}\boldsymbol{b}_{-k}(x_i))\Big)^2}b_k^2(x_i) \right)\leq 0\ \text{for}\ k=1,\dots,K, \nonumber 
\end{eqnarray}

\vspace{0.2cm}

\noindent and by Proposition 2, the conditional posteriors $p(\theta_k \vert \btheta_{-k}, \lambda, \delta, \mathcal{D}),\ k=1,\dots,K$ are strictly log-concave. As such, the GSBPS algorithm is implemented without the Griddy-Gibbs sampling step to obtain a sample from the joint posterior $p(\bfomega \vert \mathcal{D})$. The Markov chain $S_{\bfomega}$ generated by the GSBPS algorithm can be used to obtain a point estimate of $\widehat{\btheta}$ and fit a curve for the probability of success as a function of $x$, that is, $\widehat{\pi}(x)=\exp(\widehat{\btheta}^{\top}\boldsymbol{b}(x))/(1+\exp(\widehat{\btheta}^{\top}\boldsymbol{b}(x)))$, with $x\in \mathcal{B}$. Note that a $95\%$ credible interval for $\pi$ at a given value of $x$ can also be computed from $S_{\bfomega}$.\\
\indent Using our methodology, we analyze a dataset on Trypanosome organisms from a dosage-response analysis \citep{ashford1972quantal} obtained from the \textit{flexmix} R package. This dataset was already analyzed using P-splines in \cite{eilers1996flexible}. Another dataset involving Hepatitis B antibodies from \cite{eilers2021practical} is also analyzed. For the priors, we use $a_{\delta}=b_{\delta}=10^{-4}$ and $\nu=2$ and generate a Markov chain of length $M=15 000$ from which we omit the first $5000$ iterations as burn-in. The penalty parameter is initialized at $\lambda^{(0)}=1$ and the initial value of $\btheta$ is taken to be $\btheta^{(0)}=(B^{\top}B)^{-1}B^{\top}((y_1+1)/(m_1-y_1+1),\dots,(y_n+1)/(m_n-y_n+1))$.

\subsubsection{Trypanosome data}

\begin{figure}[h!]
	\begin{center}
		\includegraphics[width=13.3cm, height=6cm]{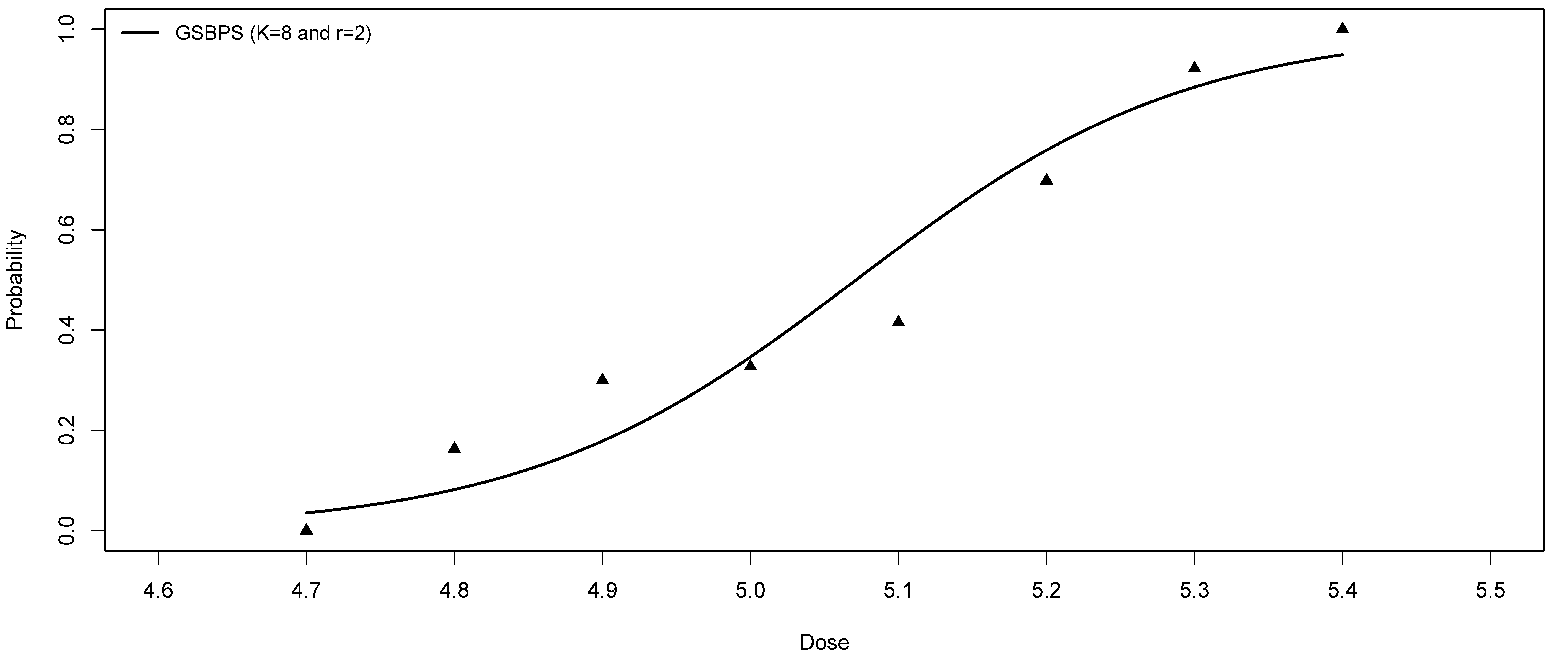}	
		\caption{Logistic regression for the Trypanosome data with $K=8$ B-splines and $r=2$ using the GSBPS procedure. The triangles show the values of the raw proportions $y_i/m_i$.}
		\label{fig3}
	\end{center}
\end{figure}

\noindent The dataset contains $n=8$ triplets $\mathcal{D}=\big((y_1,m_1,x_1),\dots,(y_8,m_8,x_8)\big)^{\top}$, where $y_i$ represents the number of Trypanosome organisms that died (out of $m_i$) after being exposed to a certain dose $x_i$. We specify $K=8$ (cubic) B-spline basis functions in the observed dose range $\mathcal{X}=[4.7,5.4]$ and a second-order penalty $r=2$. The resulting smooth fit of $\widehat{\pi}(x)$ for $x\in \mathcal{X}$ is shown in Figure~\ref{fig3}. The Markov chain generated by the GSBPS algorithm provides a fit that is closely capturing the nonlinear trend of the raw observations $y_i/m_i$ (black triangles). Furthermore, note that the algorithm works without flaws even if the data has a small sample size.

\vspace{0.1cm}

\vspace{-1.1cm} \subsubsection{Hepatitis B data}

\begin{figure}[h!]
	\begin{center}
		\includegraphics[width=12.8cm, height=5.5cm]{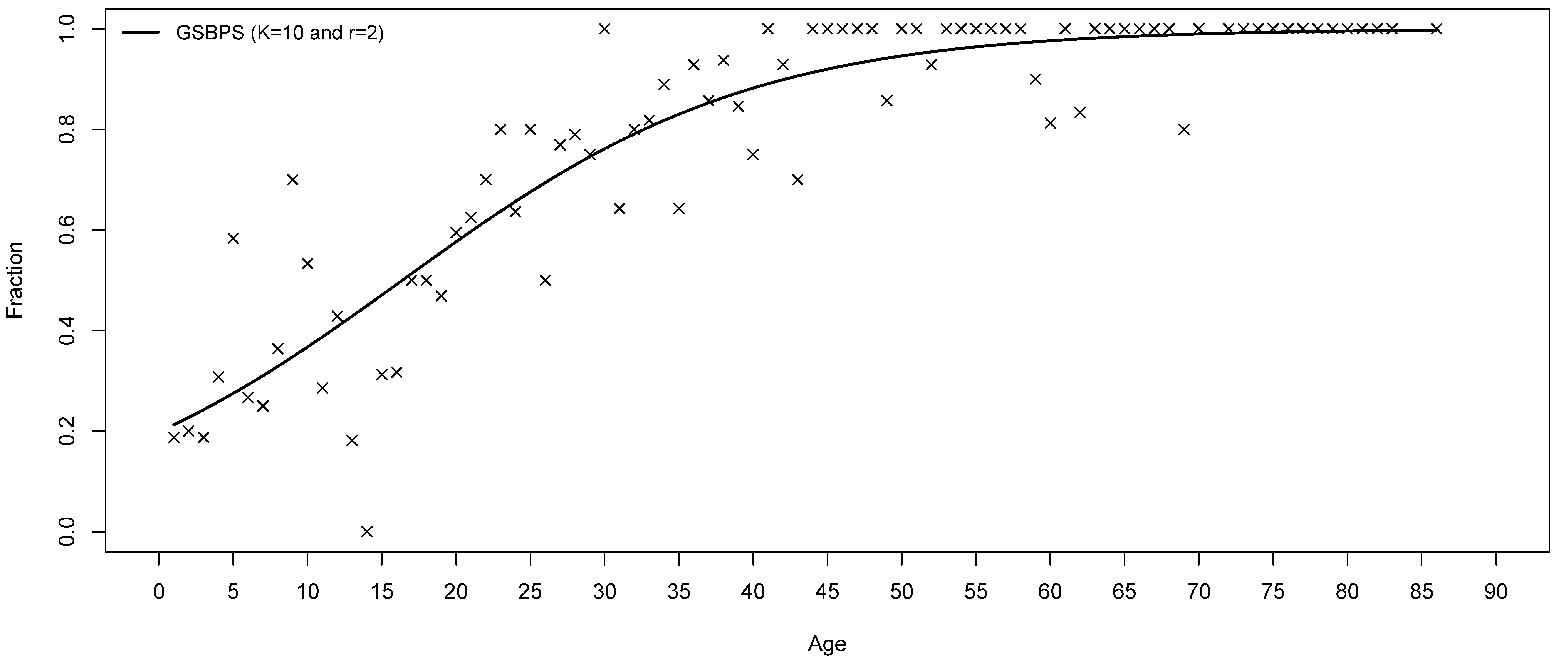}	
		\caption{Logistic regression for the Hepatitis B data with $K=10$ B-splines and $r=2$ using the GSBPS algorithm. Crosses show the values of the raw proportions $y_i/m_i$.}
		\label{fig4}
	\end{center}
\end{figure}

\noindent This dataset was obtained from the \textit{JOPS} package and contains observations on prevalence of Hepatitis B among Bulgarian males (see \cite{keiding1991age} for the original data source). Observations are organized in $n=86$ triplets $\mathcal{D}=\big((y_1,m_1,x_1),\dots,(y_{86},m_{86},x_{86})\big)^{\top}$, where $x_i$ is a covariate representing \textit{Age} and $y_i$ is the number of infected persons out of $m_i$ sampled persons. The GSBPS algorithm is implemented with $K=10$ (cubic) B-spline basis functions in $\mathcal{X}=[1,86]$ and $r=2$. The fit shown in Figure~\ref{fig4} captures the nonlinear patterns of the raw fractions given by $y_i/m_i$.

\vspace{0.1cm}

\newpage 

\subsection{Negative binomial regression}

\noindent Working with a Poisson distribution on count data implies the assumption that the (population) mean and variance are equal. This assumption can be relaxed by using a negative binomial distribution. Let $Y_i \sim \text{NegBin}(\mu(x_i), \rho)$ with mean $\mu(x_i)>0$, covariate $x_i \in \mathbb{R}$ and overdispersion parameter $\rho>0$. The probability mass function is parameterized so that $\mathbb{E}(Y_i)=\mu(x_i)$ and $\mathbb{V}(Y_i)=\mu(x_i)+\mu(x_i)^2/\rho$. Moreover, denote by $\mathcal{D}=(y_1,\dots,y_n)^{\top}$ a sample of $n$ counts assumed to be independent. The latter counts can be smoothed by modeling the (log) mean as a linear combination of B-splines $\log(\mu(x_i))=\btheta^{\top}\boldsymbol{b}(x_i)$, where $\btheta = (\theta_1,\dots,\theta_K)^{\top} \in \mathbb{R}^K$ denotes the vector of B-spline coefficients and $\boldsymbol{b}(\cdot)=(b_1(\cdot),\dots,b_K(\cdot))^{\top}$ is a (cubic) B-spline basis defined on a compact support $\mathcal{B}\supseteq \mathcal{X}$, where $\mathcal{X}$ represents the data range for the covariate. This negative binomial model plays a key role in the Epidemiological modeling with Laplacian-P-Splines (EpiLPS) framework developed by \cite{gressani2022epilps}. In EpiLPS, the epidemic curve - a time series of counts representing for instance the daily number of hospitalizations or number of cases by date of symptom onset - is smoothed with P-splines. The smoothed epidemic curve is then used in a renewal equation model to obtain estimates of the time-varying reproduction number (a key quantity in infectious disease models to characterize the transmissibility of a pathogen). In EpiLPS, estimates of spline coefficients can be obtained either through a sampling-free approach relying on the Laplace approximation or alternatively through a fully Markov chain Monte Carlo sampler. The MCMC version of EpiLPS uses the Metropolis-adjusted Langevin algorithm (MALA) \citep{roberts1996exponential} and the EpiLPS-MALA method has recently been shown to provide robust uncertainty quantification of the time-varying reproduction number \citep{steyn2025}. The MALA algorithm provides a solid and reliable workhorse to sample the joint posterior distribution in the EpiLPS model, yet its functioning is much harder to understand as compared to a more classic Gibbs sampling scheme. We therefore propose to illustrate that our GSBPS algorithm is also a viable option in a negative binomial regression setting as encountered for instance in the EpiLPS framework. In the current negative binomial regression setting, the vector of all model parameters is  $\bfomega=(\btheta^{\top}, \bfeta^{\top})^{\top}$, with $\bfeta^{\top}=(\lambda, \delta, \rho)$. The conditional log-likelihood function of $\theta_k$ for $k=1,\dots,K$ satisfies:

\vspace{-0.4cm}

\begin{eqnarray}
	\frac{\partial^2 \ell(\theta_k; \btheta_{-k}, \rho, \mathcal{D})}{\partial \theta_k^2}=-\rho\sum_{i=1}^n \left(\frac{(y_i+\rho)\exp(-\theta_k b_k(x_i))\exp(-\btheta_{-k}^{\top}\boldsymbol{b}_{-k}(x_i))}{\Big(\rho+\exp(-\theta_k b_k(x_i))\exp(-\btheta_{-k}^{\top}\boldsymbol{b}_{-k}(x_i))\Big)^2}b_k^2(x_i) \right)\leq 0, \nonumber 
\end{eqnarray}

\vspace{0.2cm}

\noindent and by Proposition 2, the conditional posteriors $p(\theta_k \vert \btheta_{-k}, \lambda, \delta, \rho, \mathcal{D}),\ k=1,\dots,K$ are strictly log-concave. In EpiLPS, a Gamma prior is imposed on the overdispersion parameter $\rho \sim \mathcal{G}(a_{\rho},b_{\rho})$, yielding the following conditional posterior for $\breve{\rho}=\log(\rho)$:

\vspace{-0.4cm}

\begin{eqnarray}
p(\breve{\rho} \vert \btheta, \lambda, \delta, \mathcal{D}) \propto \mathcal{L}(\btheta, \breve{\rho}) \exp(a_{\rho}\breve{\rho}-b_{\rho}\exp(\breve{\rho})). \nonumber
\end{eqnarray}

\noindent Showing that $p(\breve{\rho} \vert \btheta, \lambda, \delta, \mathcal{D})$ is log-concave is not an easy task and we therefore decide to use the Griddy-Gibbs sampler to explore the latter conditional posterior in the GSBPS algorithm. To have a fair comparison, we use the same prior structure as in EpiLPS, namely $a_{\delta}=b_{\delta}=10$, $\nu=2$ and $a_{\rho}=b_{\rho}=10^{-4}$. We initialize our GSBPS algorithm at $\lambda^{(0)}=\rho^{(0)}=1$ and use the same starting value for $\btheta$ as in the Poisson model (cf.\ Section 4.1). The GSBPS algorithm is applied to daily incidence data of the Zika virus in Girardot, Colombia, reported by \cite{rojas2016epidemiology}. The dataset is available in the R package \textit{outbreaks}. We use the GSBPS algorithm as well as the EpiLPS-MALA algorithm from the EpiLPS package to generate two Markov chains of length $M=5 000$ in order to compare the smoothed epidemic curves.

\subsubsection{Zika virus data}

\begin{figure}[h!]
	\begin{center}
		\includegraphics[width=16cm, height=7.2cm]{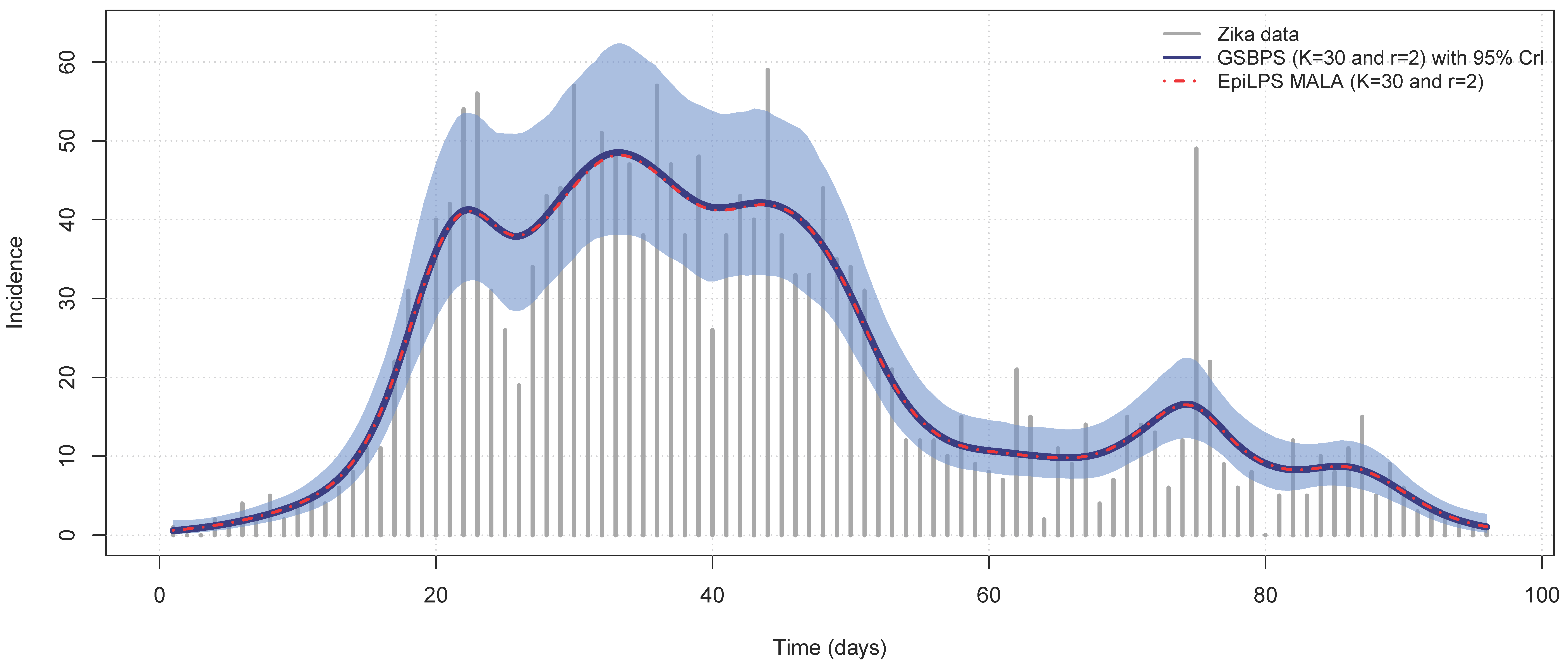}	
		\caption{Epidemic curve for the Zika virus data (vertical bars), smoothed epidemic curve with EpiLPS-MALA (dashed) and with the GSBPS algorithm (solid). The shaded surface represents a $95\%$ credible interval (CrI) for the negative binomial mean $\mu$ obtained with the GSBPS procedure.}
		\label{fig5}
	\end{center}
\end{figure}

\noindent The incidence data spans the period October 2015 - January 2016 and an epidemic curve can be constructed over $96$ days. The GSBPS and EpiLPS-MALA algorithms are implemented with $K=30$ B-splines in $\mathcal{X}=[1, 96]$ and a second-order penalty $r=2$. Figure~\ref{fig5} shows the epidemic curve (vertical bars) and the smoothed epidemic curve obtained with EpiLPS-MALA (dashed). The smoothed epidemic curve obtained with the GSBPS algorithm (solid) is hard to distinguish from the dashed curve, showing that cycling through the conditional posteriors is also a good strategy to explore the joint posterior distribution in EpiLPS. This encouraging result shows the strong potential of the GSBPS procedure. Finally, it is worth mentioning that trace plots of the logposterior and the hyperparameters $\lambda$, $\delta$, and $\rho$ obtained with the GSBPS algorithm (available here \url{https://github.com/oswaldogressani/GSBPS}) show good mixing behavior, reflecting efficient posterior exploration.

\vspace{0.1cm}

\vspace{-0.6cm}\section{Conclusion}

\noindent MCMC procedures in Bayesian P-splines models are usually based on Metropolis-type algorithms that require a prudently chosen proposal distribution for efficient exploration of the posterior. To circumvent the delicate task of selecting an appropriate proposal distribution, we orient the sampling strategy towards a Gibbsian scheme. Exploring the unique characteristics of conditional posteriors in Bayesian P-splines models, we propose a Gibbs mechanism that alternates between adaptive rejection sampling and Griddy-Gibbs sampling depending on whether or not the target conditional has a log concave shape. Our approach is calibration-free in the sense that it avoids the choice and tuning of a proposal distribution encountered in classic Metropolis-Hastings algorithms and is therefore an appealing tool for users with little statistical expertise in need of a more automatic way to carry out Bayesian inference with MCMC. The proposed GSBPS algorithm is relatively easy to understand and can be translated in a compact and user-friendly R routine. This makes our approach ready to use for practical problems covering a wide range of statistical applications as shown in this paper for density estimation and (negative) binomial regression. The proposed GSBPS algorithm has many attractive features described in the following contexts.

\noindent \underline{Verify the accuracy of Laplace approximations}\\
\noindent Laplacian-P-splines rely on the Laplace approximation to posterior distributions and open the door to a sampling-free method for inference. The main benefit is a drastic gain in computational performance as compared to MCMC algorithms. The GSBPS procedure offers an automatic toolbox that can be used to assess the accuracy of the Laplace approximation by comparing it against the distribution of the MCMC sample obtained with the GSBPS algorithm. When the MCMC sample reveals a skewed distribution, replacing the Laplace approximation with another approximation based on a skewed target may be an interesting option \citep[see][]{lambert2023penalty}.\\

\noindent \underline{Provide initial conditions for Laplace approximations}\\
\noindent The Laplace approximation in Bayesian P-splines models is usually implemented on a multidimensional target posterior around its modal value. The mode is typically obtained via optimization algorithms (e.g.\ Newton-Raphson or Levenberg-Marquardt) that have to be initialized by specifying a starting value in the parameter space. Initialization with uniformative starting points may lead to divergence issues. The GSBPS algorithm could be used to ``learn'' about the posterior parameter space and provide a smart guess of a suitable starting point for computation of the Laplace approximation.\\

\noindent \underline{Natural quantification of uncertainty}\\
\noindent The GSBPS methodology is a MCMC method that generates a sample from the joint posterior involving all model parameters. As such, it permits to straightforwardly construct a credible interval for any model parameter (and in particular the hyperparameters) or any function thereof.\\

\noindent From here, several research directions can be explored. Currently, the GSBPS algorithm is coded in pure R language. For the analyses carried out in this paper, a few minutes were required to explore the posterior. To be computationally more efficient, it would be worth coding the loop in the GSBPS algorithm in C\texttt{++}. Furthermore, the Griddy-Gibbs algorithm used here is based on a simple grid construction with equidistant points. More sophisticated designs are possible. For instance, constructing a grid with a more dense concentration of points in regions of high mass and a less dense concentration of points in regions of low mass \citep{ritter1992facilitating} may be a more efficient strategy. In the Griddy-Gibbs step of our algorithm, it could also be interesting to use a more clever approximations to the cumulative distribution function (e.g.\ a piecewise linear or piecewise quadratic approximation) as compared to the current piecewise constant approximation. Finally, it would be interesting to assess how the proposed GSBPS algorithm performs in complex models characterized by more challenging likelihood functions, high dimensional parameter spaces or different penalty structures.

\vspace{-0.45cm}

\section*{Data availability}
Simulation results and real data applications in this paper can be fully reproduced with the code available on GitHub \url{https://github.com/oswaldogressani/GSBPS}.

\vspace{-0.25cm}

\section*{Competing interests}
The authors have declared that no competing interests exist.

\vspace{-0.15cm}

\bibliographystyle{apa}
\bibliography{Bibliography}
	
\section*{Appendix A1: Proof of Proposition 1} 

\noindent Denote by $p_{jk}$ the $(j,k)$ entry of the $K \times K$ symmetric penalty matrix $P$ constructed from a penalty order $r$. The log prior of the global smoothness prior $(\btheta \vert \lambda)\sim \mathcal{N}(0,(\lambda P)^{-1})$ is given by:

\vspace{-0.5cm}

\begin{eqnarray}
\log p(\btheta \vert \lambda)&\dot{=}&0.5K\log(\lambda)-0.5\lambda \btheta^{\top}P\btheta \nonumber \\
&\dot{=}& 0.5K\log(\lambda)-0.5\lambda\left(\sum_{k=1}^K p_{kk}\theta_k^2+\sum_{k=1}^K \sum_{\substack{j=1 \\ j\neq k}}^K p_{kj}\theta_k \theta_j\right), \nonumber 
\end{eqnarray}

\vspace{0.1cm}

\noindent where $\dot{=}$ denotes equality up to an additive constant. The conditional prior mean of the $k$th B-spline coefficient $\mathbb{E}(\theta_k \vert \btheta_{-k}, \lambda)$ is obtained by solving the linear system $\partial \log p(\btheta \vert \lambda)/\partial \theta_k=0$ for $\theta_k$ and the conditional prior variance is $\mathbb{V}(\theta_k \vert \btheta_{-k},\lambda)=(-\partial^2 \log p(\btheta \vert \lambda)/\partial \theta_k^2)^{-1}$. Note that:

\vspace{-0.2cm}

\begin{eqnarray}
\frac{\partial \log p(\btheta \vert \lambda)}{\partial \theta_k}=-0.5\lambda \left(2p_{kk}\theta_k+2\sum_{\substack{j=1 \\ j\neq k}}^K p_{kj} \theta_j\right)
=-\lambda \left(p_{kk}\theta_k+\sum_{\substack{j=1 \\ j\neq k}}^K p_{kj} \theta_j\right). \nonumber 
\end{eqnarray}

\vspace{0.3cm}

\noindent Solving $\partial \log p(\btheta \vert \lambda)/\partial \theta_k=0$ for $\theta_k$ yields the conditional prior mean:

\vspace{-0.3cm}

\begin{eqnarray} 
\mathbb{E}(\theta_k \vert \btheta_{-k}, \lambda)=-(p_{kk})^{-1} \sum_{\substack{j=1 \\ j\neq k}}^K p_{kj} \theta_j. \nonumber 
\end{eqnarray}

\vspace{-0.1cm}

\noindent The conditional variance is simply:

\begin{eqnarray} 
\mathbb{V}(\theta_k \vert \btheta_{-k},\lambda)=(-\partial^2 \log p(\btheta \vert \lambda)/\partial \theta_k^2)^{-1}=(\lambda p_{kk})^{-1}. \nonumber 
\end{eqnarray}

\vspace{0.2cm}

\noindent The diagonal entries $p_{kk}$ depend on $\varepsilon$ (since $D_r^{\top}D_r$ is perturbed on the main diagonal by $\epsilon$ to satisfy full rankedness) and also on the chosen penalty order $r$. To make this dependence explicit, we write $z_r(k, \varepsilon):=p_{kk}$. The terms involved in the sum of the conditional mean formula will depend on $r$ and  $\btheta_{-k}$, and we use the notation $\psi_r(\btheta_{-k}):=-\sum_{\substack{j=1 \\ j\neq k}}^K p_{kj} \theta_j$ to make this explicit. Using the latter notations, the conditional mean is written as $\mathbb{E}(\theta_k \vert \btheta_{-k}, \lambda)=\psi_r(\btheta_{-k})z_r^{-1}(k,\varepsilon)$ and the conditional variance is $\mathbb{V}(\theta_k \vert \btheta_{-k},\lambda)=(\lambda z_r(k,\varepsilon))^{-1}$. To summarize, for a penalty of order $r$, the conditional priors of the B-spline coefficients are:

\vspace{-0.2cm}

\begin{eqnarray}
	p(\theta_k \vert \btheta_{-k}, \lambda)&=&p_G\Big(\theta_k;\psi_r(\btheta_{-k})(z_r(k, \varepsilon))^{-1},(\lambda z_r(k, \varepsilon))^{-1}\Big) \nonumber \\
	&=& \left(\frac{\lambda z_r(k, \varepsilon)}{2 \pi}\right)^{0.5} \exp\Bigg(-\frac{\lambda z_r(k, \varepsilon)}{2}\Bigg(\theta_k-\frac{\psi_r(\btheta_{-k})}{z_r(k, \varepsilon)}\Bigg)^2\Bigg),\ \ k=1,\dots,K\ \  \square \nonumber 
\end{eqnarray}

\end{document}